%% file: main.tex
\documentclass{PBCreport}

\IfFileExists{images_cls/PBC_logo2022.png}{}{}

\usepackage{wrapfig}
\usepackage{comment}
\usepackage{booktabs}
\usepackage{afterpage}
\usepackage[T1]{fontenc}
\usepackage{textcomp}
\usepackage{color, colortbl}
\usepackage{booktabs}
\usepackage{afterpage}
\usepackage{tabularx}
\usepackage{siunitx}
\usepackage{hyperref}
\usepackage{array}
\usepackage{caption}
\usepackage{geometry}
\usepackage{longtable}
\usepackage{float}
\usepackage{graphicx,mathptmx,amsmath,amsfonts,amscd,amsthm,amssymb,eucal,psfrag,color,subfig,url,multirow,gensymb}
\usepackage{rotating}
\usepackage{soul}
\usepackage{appendix}
\usepackage{placeins}
\usepackage{subcaption}
\hypersetup{
     colorlinks = true,
     linkcolor = blue,
     anchorcolor = blue,
     citecolor = blue,
     filecolor = blue,
     urlcolor = blue
     }

\definecolor{LightCyan}{rgb}{0.88,1,1}
\definecolor{LightYellow}{rgb}{1,0.97,0.9}

\documentlabel{CERN-PBC Report-2025-004}

\begin{document}

\title{A Long-Baseline Atom Interferometer at CERN LHC Point 4: Implementation Study}

\author{\parbox{\textwidth}{\it 
G.~Arduini$^{1}$, 
O.~Buchmueller$^{2,3}$, T.A.~Bud$^1$,
S.~Calatroni$^{1,*}$, O.~Crespo-Lopez$^1$, A.~Devienne$^1$, J.~Ellis$^{1,4}$, T.~Hakulinen$^1$, A.~Infantino$^1$, D.~Lafarge$^1$, A.P.~Marion$^1$}\\
~~\\
%\email{author.email@cern.ch}
%\affiliation{CERN, CH 1211 Geneva 23, Switzerland}
\small{$^1$ CERN, 
$^2$ Imperial College London, $^3$ University of Oxford, $^4$ King's College London}\\
\small{$^*$ Editor}}

%\newpage

\abstract{
Building on the feasibility study in \cite{Arduini:2851946}, this report supported by the Physics Beyond Colliders (PBC) Study Group describes the technical implementation of modifications to the PX46 shaft at LHC Point 4 during LS3 (June 2026 – June 2030) that would enable it to accommodate the installation and operation of a vertical long-baseline Atom Interferometer during Run 4 without affecting LHC operations. We specify in detail the necessary civil‐engineering work, installation of bespoke radiation shielding, deployment of access‐control systems and safety alarms, and design of a mobile elevator platform. Our comprehensive technical assessment identifies no fundamental obstacles or showstoppers to implementation. Refined cost estimates and a critical‐path schedule confirm that, from formal approval, all interventions can be completed within a 1.5-year window. These preparations would ensure seamless, concurrent operation of the Atom Interferometer experiment and the HL-LHC, with all technical challenges successfully addressed through established engineering solutions.
}

\maketitle

\tableofcontents

\newpage

%\Section{Introduction (G. Arduini, S. Calatroni, O. Buchmuller, J. Ellis)}
\input{1-Introduction.tex}

%\Section{Executive summary (G. Arduini, S. Calatroni, O. Buchmuller, J. Ellis)}
\input{2-ExecutiveSummary.tex}

%\Section{Physics motivations (O. Buchmuller, J. Ellis)}
\input{3-PhysicsMotivations.tex}

%\Section{Summary of the feasibility study (S. Calatroni)}
\input{4-FeasibilitySummary.tex}

%\section{Civil engineering - shielding wall (T.A. Bud)}
\input{5-CivilEngineering.tex}

%\subsection{Access control and safety systems (T. Hakulinen)}
\input{6-AccessControl.tex}

%\subsubsection{Lifting platform technical requirements (D. Lafarge)}
\input{7-LiftingPlatform.tex}
\input{8-HVAC.tex}

%\subsubsection{Electricity (M. Parodi)}
\input{9-OtherServices.tex}

%\section{Cost and Schedule (All)}
\input{10-CostSchedule.tex}
\input{11-Conclusions.tex}

\vspace{0.5cm}
\paragraph{Acknowledgements}

The authors would like to thank the following colleagues for their contributions and support at different stages of this study: M.~Barberan~Marin, M.~Bernardini, O.~Boettcher, N.~Broca, F.~Corsanego, P.~Durand, Ch.~Gaignant, F.~Gerigk, G.~Godineau, M.~Guinchard, M.~Majstorovic, A.~Navascues~Cornago, M.~Parodi, F.~Pillon, J.~Ramos, V.~Rios, J.-Ph.~Tock, K.~Turaj, M.~Van~de~Veire.

%\textbf{\color{red} Others to be acknowledged?}

\bigskip

%This is an example of citation \cite{Gorzawski:1697022}
%\clearpage
%\section*{Acknowledgments}
%Place for acknowledgements.\\
%The work of J.E. was supported in part by the United Kingdom STFC Grants
%ST/T00679X/1 and ST/T000759/1.

%\begin{thebibliography}{90}

%\bibitem{review} A reference

%\end{thebibliography}
\newpage

\bibliography{PBCreport}

\newpage
\appendix
\section{Definitions of acronyms}
\label{sec:Acronyms}

\textbf{}
\setlength{\parindent}{0pt}

\noindent
\textbf{AI}: Atom Interferometer/Interferometry

\textbf{AION}: Atom Interferometer Observatory and Network

\textbf{ASN}: Atom Shot Noise

\textbf{AURIGA}: Antenna Ultracriogenica Risonante per l'Indagine Gravitazionale Astronomica~\footnote{Ultracryogenic Resonant Bar Gravitational Wave Detector}

\textbf{BH}: Black Hole

\textbf{BSM}: Beyond the Standard Model

\textbf{CV}: Cooling and Ventilation

\textbf{DM}: Dark Matter

\textbf{EM}: ElectroMagnetic

\textbf{EMC}: Electromagnetic Compatibility

\textbf{GGN}: Gravity Gradient Noise

\textbf{GW}: Gravitational Wave

\textbf{HL-LHC}: High-Luminosity LHC

\textbf{HVAC}; Heating, Ventilation and Air Conditioning

\textbf{IMBH}: Intermediate-Mass Black Holes

\textbf{KAGRA}: KAmioka GRAvitational wave detector

\textbf{LACS}: LHC Access Control System

\textbf{LASS}: LHC Access Safety System

\textbf{LHC}: Large Hadron Collider

\textbf{LIGO}: Laser Interferometer Gravitational Observatory experiment

\textbf{LISA}: Laser Interferometer Space Antenna

\textbf{LMT}: Large Momentum Transfer

\textbf{LS}: Long Shutdown

\textbf{LSBB}: Laboratoire Souterrain {\` a} Bas Bruit~\footnote{Low-Noise Underground Laboratory}

\textbf{LVK}: LIGO, Virgo and KAGRA

\textbf{MAD}: Material Access Device

\textbf{MAGIS}: Matter-wave Atomic Gradiometer Interferometric Sensor experiment

\textbf{MCI}: Maximum Credible Incident

\textbf{MICROSCOPE}: Micro-Satellite à traînée Compensée pour l'Observation du Principe d'Equivalence~\footnote{Micro-Satellite with Compensated Drag for Observing the Principle of Equivalence}

\textbf{MIGA}: Matter wave-laser based Interferometer Gravitation Antenna

\textbf{NHNM}: New High-Noise Model

\textbf{NLNM}: New Low-Noise Model

\textbf{ODH}: Oxygen Deficiency Hazard

\textbf{PAD}: Personal Access Device

\textbf{PBC}: Physics Beyond Colliders

\textbf{PM}: Puit Materiel~\footnote{Access shaft with stairs and lift used for the transfer of equipment}

\textbf{PPE}: Personal Protection Equipment

\textbf{PX}: Puit eXperience~\footnote{Access shaft to experimental cavern for (formerly) LEP or (currently) LHC detectors}

\textbf{PX46}: Access shaft at LHC Point 4

\textbf{RP}: Radiation Protection

\textbf{RF}: Radio Frequency

\textbf{SMBH}: Super Massive Black Holes

\textbf{SM}: Standard Model

\textbf{SUSI}: système de SUrveillance des SItes~\footnote{Site Surveillance system}

\textbf{SU4}: Surface building dedicated to the cooling and ventilation at Point 4

\textbf{SX4}: Surface building on top of the PX46 shaft

\textbf{TETRA}: Terrestrial Trunked Radio, formerly known as Trans-European Trunked Radio

\textbf{TVLBAI}: Terrestrial Very-Long-Baseline Atom Interferometer

\textbf{TX46}: Access gallery at LHC Point 4

\textbf{ULDM}: Ultra-Light Dark Matter

\textbf{UX45}: Experimental cavern at LHC Point 4

\textbf{WIMP}: Weakly Interacting Massive Particle

\textbf{YETS}: Year-End Technical Stop

\textbf{ZAIGA}: Zhaoshan Long-baseline Atom Interferometer Gravitation Antenna
\end{document}

%% file: 1-Introduction.tex
\section{Introduction}
%(G. Arduini, S. Calatroni, O. Buchmuller, J. Ellis)
\label{sec:Intro}

Atom interferometer (AI) experiments use quantum sensing technology to search for
the interactions of ultralight bosonic dark matter with
Standard Model (SM) particles, conduct a pioneering search for gravitational waves in an unexplored frequency band, and make other unique fundamental physics measurements.
Vertical shafts of $\sim 100$~m, such as those providing access to the Large Hadron Collider (LHC), are promising locations for the next generation of such experiments.
Exploratory studies of the PX46 LHC access shaft supported by the Physics Beyond Colliders (PBC) Study Group at CERN have shown that {\it prima facie} it is a suitable
location for an atom interferometer experiment such as the AION-100 proposal~\cite{Badurina:2019hst}. 
This exploratory study 
analysed the principal technical requirements for such an AI experiment and their implications, 
identifying potential engineering solutions and discussing principal cost and schedule drivers~\cite{Arduini:2851946}. Here we analyse in more detail the technical constraints, the time schedule and the cost estimates for adapting the PX46 shaft and its surrounding infrastructure during the LHC third Long Shutdown (LS3) \cite{LTS:ACC-PM-MS} in preparation for the installation of an AI during the LHC Run 4, in parallel with High-Luminosity LHC (HL-LHC) machine operation. 

Explanations of the acronyms used in this document are listed in the Appendix~\ref{sec:Acronyms}.

%% file: 2-ExecutiveSummary.tex
\section{Executive summary}
%(G. Arduini, S. Calatroni, O. Buchmuller, J. Ellis)
\label{sec:Exec}

AIs offer interesting prospects for for searches for Dark Matter (DM) and Gravitational Waves (GWs)~\cite{Dimopoulos:2008sv} that are largely complementary to established techniques.
%Quantum sensors are attracting increasing attention for their potential to make precise measurements within the Standard Model (SM) and search for possible new physics beyond the Standard Model (BSM). Among the proposed applications of Quantum Technology to Fundamental Physics, one of the most interesting is Atom Interferometry (AI), which offers interesting prospects for searches for Dark Matter (DM) and Gravitational Waves (GWs)~\cite{Dimopoulos:2008sv} that are largely complementary to established techniques.
They exploit features of state-of-the-art atomic clocks together with techniques used for building inertial sensors. There are several AI experiments and projects using baselines that are ${\cal O}(10)$~m, sited in university laboratories or research institutes (e.g., at Stanford~\cite{Overstreet:2021hea}, Oxford~\cite{Badurina:2019hst,Bongs:2025rqe}), Hannover~\cite{schlippert2020matter} and Wuhan~\cite{zhou2011development}). In the long run, the Terrestrial Very Long Baseline Atom Interferometer (TVLBAI) Proto-Collaboration envisages one or more networked experiments with baselines that are ${\cal O}(1)$~km, as detailed in the summary reports~\cite{abend_terrestrial_2024, abdalla_terrestrial_2025}. Essential intermediate steps towards this ambitious goal are experiments with baselines that are ${\cal O}(100)$~m (e.g.,  at Fermilab~\cite{MAGIS-100:2021etm} and in France~\cite{Canuel:2017rrp}.

\begin{figure}[!]
%\centering
%\begin{wrapfigure}[41]{lt}{0.5\textwidth}
%\begin{figure}[t!]
\centering
%~~\\
%\vspace{-0.2cm}
\includegraphics[width=1.0\textwidth]{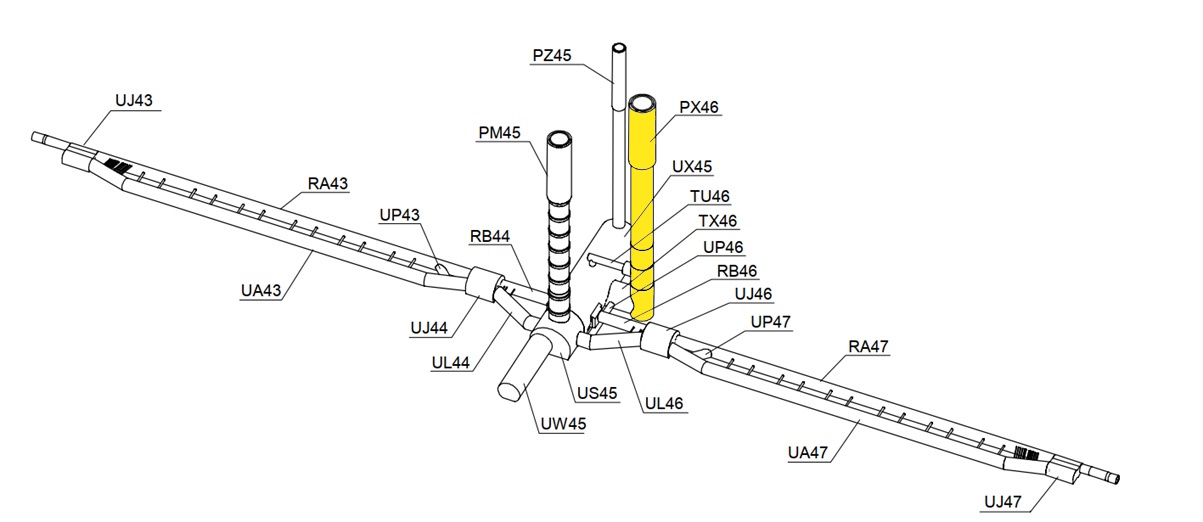}
%\hspace{5mm}
%\includegraphics[width=0.45\textwidth]{Figures//Feasibility_Study/3-PhysicsMotivations/Experiment.png}
%\quad
%\includegraphics[width=0.45\textwidth]{Figures//Feasibility_Study/3-PhysicsMotivations/PX46Layout.png}
%\includegraphics[width=0.312\textwidth]{AIONSchem.pdf}
\caption{
% {\it Top panel}:
Schematic drawing of the civil engineering infrastructure at Point~4 on the LHC ring,
showing the LHC tunnel, the PX46 shaft highlighted in yellow, the UX45 cavern where the LHC RF power system is located, and the horizontal TX46 gallery. LHC machine components are installed in the adjacent tunnel sections RA43, RB44, RB46 and RA47.
%{\it Bottom left panel}: Horizontal cross section of the PX46 shaft.
%{\it Bottom right panel}: Illustration how a 100~m vertical AI similar to AION-100 could be accommodated in PX46. The experiment and its magnetic shielding would be contained within a cylinder with a diameter $\sim 1$~m, with provision for external access via a mobile platform with approximate dimensions 2~m$\times$5~m.
}
%For clarity, the sizes of the atom interferometers are shown on an exaggerated scale.}
\label{fig:PX46situation}
%\end{wrapfigure}
\end{figure}

CERN is an interesting possible location for an AI of length ${\cal O}(100)$~m, by virtue of its unique physical and technical infrastructure including vertical LHC access shafts, and its experience in hosting international experimental collaborations. CERN's PBC programme has recognised this potential for extending the scope of the CERN experimental programme in exciting new directions without impacting the exploitation of the LHC, and has organised a study of the feasibility of installing such an AI at CERN~\cite{Arduini:2851946}. This initial study identified the PX46 access shaft as the most promising site for a ${\cal O}(100)$~m vertical AI. Figure~\ref{fig:PX46situation} shows a schematic view of PX46 and the layout of the civil engineering infrastructure at Point~4 on the LHC ring. The vertical height from the SX4 surface building down to the level of the LHC is $\sim 143$~m and the internal diameter of the shaft is 10.1~m. PX46 provides access to the main LHC radiofrequency (RF) system and its primary use is for raising and lowering technical equipment. However, not all the horizontal cross-section of PX46 is required for this purpose, and there is ample space for accommodating a ${\cal O}(100)$~m AI experiment.

% However, as seen in the bottom left panel of Fig.~\ref{fig:PX46situation}, a substantial fraction of the horizontal cross-section of the PX46 shaft is not required for LHC access, and would be large enough to accommodate an AI experiment such as AION-100, as illustrated in the bottom right panel of Fig.~\ref{fig:PX46situation}.

We described in the previous report~\cite{Arduini:2851946} the general civil infrastructure on the surface and below ground at LHC Point 4, and outlined the considerations that led to the choice of PX46. %The PX46 shaft is currently used mainly for transport in and out of the LHC of machine equipment. 
The space constraints due to the transport of LHC machine components were taken into account in the feasibility study. 
We also reported in~\cite{Arduini:2851946} exploratory seismic measurements at the top and bottom of PX46. These were used to estimate the level of Gravity Gradient Noise (GGN), an important source of background, which was found to be similar to other prospective experimental sites. 
An issue specific to PX46 is the electromagnetic (EM) noise associated with the LHC RF system and other electrical equipment at depth and on the surface.
The EM noise levels were also measured and found not to be of concern~\cite{Arduini:2851946}. The variation in the ambient magnetic field during a ramp of the LHC magnets was monitored and found to be sufficiently small and slow to be dealt with by the magnetic shield of the AI. 

In view of the (unlikely) possibility of a catastrophic LHC beam loss near the base of the PX46 shaft, radioprotection measures have been foreseen to protect the operators~\cite{Arduini:2851946}, namely the construction of a protective shielding wall in TX46 at the base of PX46. This would require an access door interlocked with the general LHC machine access system and provision for temporary opening when LHC equipment must be installed or removed, as discussed in~\cite{elie_updated_2022}. 

\begin{figure}
    \centering
    \includegraphics[width=0.8\linewidth]{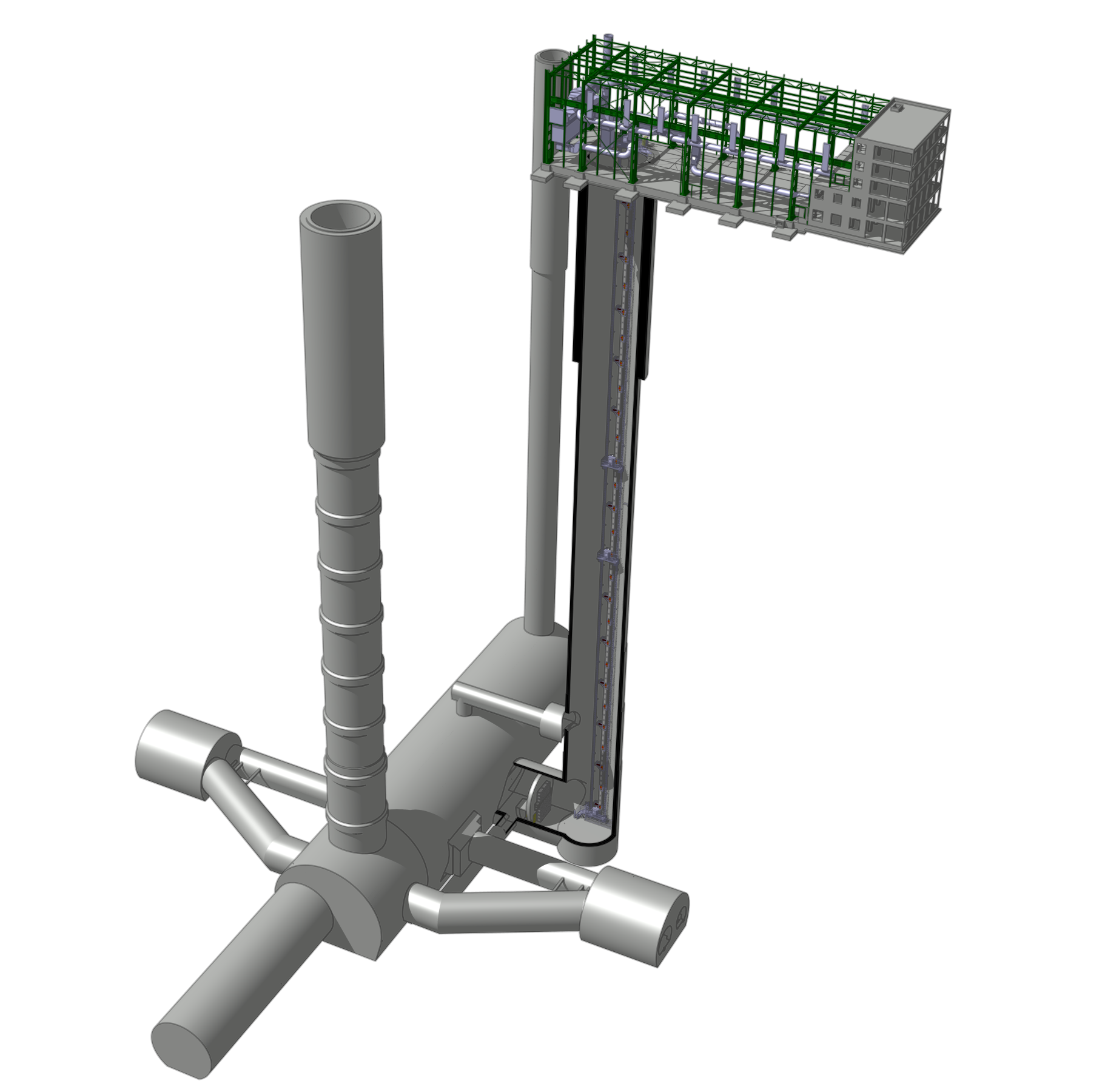}
    \caption{General schematic view of the possible implementation of an AI in the PX46 access shaft. }
    \label{fig:artistic}
\end{figure}

General safety considerations imply provisions for rapid evacuation of personnel from the experiment when necessary, such as in the event of a fire in the nearby UX45 cavern or of a major helium release from the LHC cryogenic system. This requirement implies that an {\it ad hoc} fast and secure elevator platform should be implemented, which could also serve as normal access method to the different levels of the experiment. Preliminary consultations with specialized companies reported in~\cite{Arduini:2851946} indicate that a technical solution is feasible at a reasonable cost. Suitable fire detection system, oxygen deficiency detectors and an alarm system must also be provided.

A general implementation layout of the facility, based on all the above constraints, is illustrated in Fig.~\ref{fig:artistic}, with more detailed views available in later sections of this report. We note in particular that the experimental platform requires only limited space, leaving a large fraction of the PX46 shaft cross-section available for the transport of LHC components. The transport requirement is also reflected in a large square opening in the  shielding wall, which would be closed with a movable door also made of concrete blocks that could easily be opened as needed.

These preliminary studies indicated that siting a ${\cal O}(100)$~m vertical AI in PX46 appears feasible: no showstoppers were found, and the main cost drivers for installing such an experiment were also identified~\cite{Arduini:2851946}. In particular, it has been identified that the best strategy would be to implement the essential measures to separate the PX46 shaft from the LHC and to guarantee the safety of the operators at the earliest possible occasion. Construction of an AI could then be performed at a later stage, even during LHC normal operation. A preliminary cost estimate for this was performed and found to be reasonable ($\sim$1.5~MCHF) compared to the total cost of the experiment itself (30$\div$50~MCHF). 

In the present report, after recalling the physics motivations for a ${\cal O}(100)$~m vertical AI experiment~\cite{Badurina:2019hst}, we concentrate on detailing the major interventions that will be required in order to prepare the PX46 site for the possible subsequent construction of an AI experiment, namely the construction of the shielding wall, the installation of the elevator platform and the realization of the necessary interlocked access control system. We perform an accurate study of the needed civil engineering and construction works, together with all the needed services and infrastructure. A new detailed cost evaluation is performed, resulting in a total cost estimate of the order of 1.2~MCHF, which includes all works that are essential for the purpose of separating the PX46 shaft from the LHC and all the infrastructure that would be needed for the construction of an AI experiment, but excluding any other activity that is not needed for this primary purpose.
The underlying logic is that these preparatory works should be performed at an early and convenient stage, such that the installation of the experiment could be performed in parallel with normal LHC operation. We therefore establish a technically driven schedule for our activity, resulting in about one year of preparatory studies and contract tendering, and approximately six months of field work in PX46. We analyze whether the third major LHC Long Shutdown (LS3)~\cite{LTS:ACC-PM-MS}, foreseen to allow all the hardware modifications for the High-Luminosity upgrade of the LHC (HL-LHC) and scheduled to begin in June 2026 with first beam re-injected into the LHC in July 2030, would have suitable technical windows in the planned works for achieving our goals. We find that the period from August 2027 to the end of 2028 would be well suited for the required installation activities to prepare the PX46 shaft for the subsequent safe installation and operation of an AI experiment concurrently with HL-LHC operations.

%% file: 3-PhysicsMotivations.tex
\section{Physics motivations}
%(O. Buchmuller, J. Ellis)
\label{sec:Phys}

%{\it 2-3 pages: short description of the physics goals and motivations. Why a vertical interferometer? }

The nature of Dark Matter (DM) lies beyond the scope of the Standard Model, and is one of the greatest puzzles in fundamental physics and astrophysics~\cite{Bertone:2016nfn}. Among the leading hypotheses considered are that it is composed of non-relativistic particles such as weakly-interacting massive particles (WIMPs) or coherent waves of Ultra-Light Dark Matter (ULDM) bosons. Experiments at the LHC and deep underground have not yet found any evidence for WIMPs, though searches will continue at the HL-LHC. In the meantime, there is growing interest in searches for ULDM waves, and this is one of the principal scientific objectives of AI experiments~\cite{Graham:2015ifn,Badurina:2019hst,Badurina:2021lwr}.

Another primary objective of such experiments is the search for gravitational waves (GWs)~\cite{Graham:2012sy} in a range of frequencies around 1~Hz that is intermediate between the peak sensitivities of present terrestrial laser interferometer experiments such as LIGO~\cite{LIGOScientific:2014pky}, Virgo~\cite{VIRGO:2014yos} and KAGRA~\cite{Aso:2013eba} (LVK), and the space experiment LISA~\cite{LISA:2017pwj}. Among the targets of experiments in this frequency range are mergers of black holes with masses intermediate between those with masses $\lesssim 100$ solar masses whose mergers have been detected by LIGO and Virgo and the supermassive black holes (SMBHs) with masses $\gtrsim 10^6$ solar masses detected in the centres of galaxies (see, e.g., ~\cite{EventHorizonTelescope:2019dse,EventHorizonTelescope:2022wkp}). Mergers of such intermediate-mass black holes (IMBHs) are likely to have played a key role in the assembly of SMBHs. Isolated IMBHs have been detected (see, e.g., \cite{haberle2024fast}), and the first observation of GWs from the merger of a BH with mass $> 100$ solar masses with a lower-mass BH has recently been reported by LVK~\cite{LIGOScientific:2025rsn}, but laser interferometer experiments cannot cover the intermediate frequency range where mergers of intermediate-mass BHs should be observed.

A 100~m AI would be the first to explore possible GW signals in the intermediate frequency range, paving the way for subsequent TVBAI detectors with the sensitivity to constrain models of SMBH formation (see, e.g., in~\cite{AION:2025igp}).
Detectors in this intermediate frequency range may also be sensitive to a background of GWs produced by fundamental physics processes such as the evolution of a network of cosmic strings or first-order phase transitions in the early Universe~\cite{Badurina:2019hst}. 
%The other principal objective of such experiments is the search for gravitational waves (GWs) in a range of frequencies around 1~Hz that is intermediate between the peak sensitivities of present terrestrial experiments such as LIGO~\cite{LIGOScientific:2014pky}, Virgo~\cite{VIRGO:2014yos} and KAGRA~\cite{Aso:2013eba}, and the approved space-borne experiment LISA~\cite{LISA:2017pwj}. Among the targets of experiments in this frequency range are mergers of black holes with masses intermediate between those whose mergers have been detected by LIGO and Virgo and the supermassive black holes (SMBHs) detected in the centres of galaxies~\cite{EventHorizonTelescope:2019dse,EventHorizonTelescope:2022wkp}. Mergers of intermediate-mass black holes are likely to have played a key role in the formation of SMBHs. Detectors in the intermediate frequency range may also be sensitive to a background of GWs produced by fundamental physics processes such as first-order phase transitions in the early Universe or the evolution of a network of cosmic strings~\cite{Badurina:2019hst}. 

\begin{figure}[t!]
\centering
%~~\\
\vspace{-0.7cm}
\includegraphics[width=0.43\textwidth]{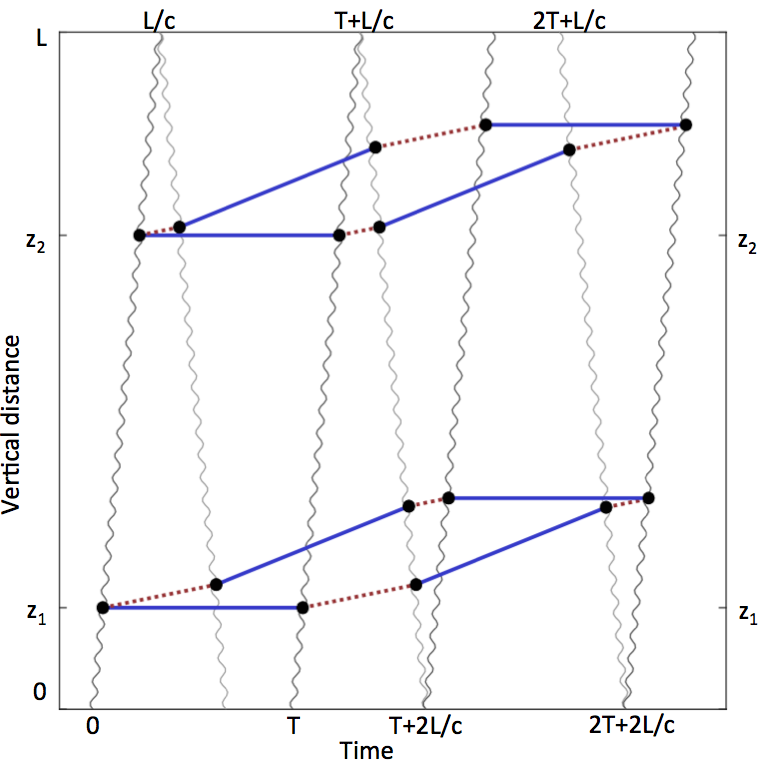}
\caption{Space-time diagram of the operation of a pair of cold-atom interferometers based on single-photon transitions between the ground state (blue) and the excited state (red dashed). Height is shown on the vertical axis and the time axis is horizontal.
The laser pulses (wavy lines) traversing the baseline from opposite ends are used to divide, redirect, and recombine the atomic matter waves, yielding interference patterns that are sensitive to the modulation of the atomic transition frequency caused by coupling to ULDM, or the space-time distortions caused by
GWs.}
\label{fig:space-time}
%\end{wrapfigure}
\end{figure}

The basic principle of an AI was described in~\cite{Arduini:2851946} and is illustrated in~Fig.~\ref{fig:space-time}. A cloud of cold atoms is split by a laser pulse into populations of ground-state and excited atoms, a second `mirror' pulse interchanges the populations of excited and ground-state atoms, which are then recombined after a flight time $\Delta T$ using another laser pulse, and the wave functions of the two atomic populations then interfere. Each laser interaction with an atom imparts momentum, so that the atom clouds follow separated trajectories, and AIs aim to increase sensitivity by using the laser to impart many momentum kicks $n$ during each cycle.  The sensitivity also increases with the AI length $L$, and the number $N$ of atoms in each cloud. The experimental cycle is repeated after a time $\Delta t$ and, to suppress the sensitivity to laser phase noise, differential measurements are made with sources separated vertically by a distance $\Delta z$ (as demonstrated, e.g., in~\cite{AION:2025igp}), over a total integration time, $T_{\rm Int}$. Indicative values of these parameters that are envisaged for an AI in PX46 are listed in Table~\ref{tab:AION100parameters}.

\begin{table}[h]
 \caption{Indicative experimental parameters for a 100~m Atom Interferometer.}
 \label{tab:AION100parameters}
  \centering
  \begin{tabular}{c|c|c|c|c|c|c|c}
   L [m] & T [s] & $n$ & $\Delta z$ [m] & $N$ & $\Delta t$ [s] & Phase noise [1/$\sqrt{\rm Hz}$] & $T_{\rm{Int}}$ [s] \\
   \hline
%   100 & 1.4 & 1000 & 85 & $10^8$ & 1.5 & $10^8$ \\
   100 & 1.4 & 5494 & 50 & $10^8$ &  1.5 & $10^{-5}$  & $10^8$
  \end{tabular} %\\
\end{table}
%%%%%
%\vspace{0.5cm}
%The conceptual scheme we consider for a vertical AI with two atom sources is shown in Figure~\ref{fig:Schematic}. 
\begin{comment}
%\begin{wrapfigure}[24]{t}{0.45\textwidth}
\begin{figure}[h!]
\centering
%~~\\
%\vspace{-0.6cm}
\includegraphics[width=0.43\textwidth]{Figures/Feasibility_Study/3-PhysicsMotivations/AIONSchem_2sources.pdf}
%\vspace{-0.4cm}
\caption{Conceptual scheme of an Atom Interferometer (AI) experiment with two atom sources that project clouds vertically, addressed by a single laser source (this diagram is not to scale).}
\label{fig:Schematic}
\end{figure}
%\end{wrapfigure}
There is a single laser source, shown here as located at the top of the vertical vacuum tube, though a location at the bottom could also be considered. One of the atom sources is located at the bottom of the vacuum tube, whereas the location of the upper source is only indicative. This upper atom source is labelled by (1), the trajectories of the atoms it launches are labelled by (2,3), and the detector to measure interference patterns is labelled by (4). Additional sources and detectors may be added as desired. We note that the vacuum pipe is surrounded by a magnetic shield.
\end{comment}
%%%%%%%%%%%%%%%%%%%%%%%%%%%%%
\vspace{3mm}
\begin{figure*}[h!]
 \centering
 \includegraphics[width=0.48\textwidth]{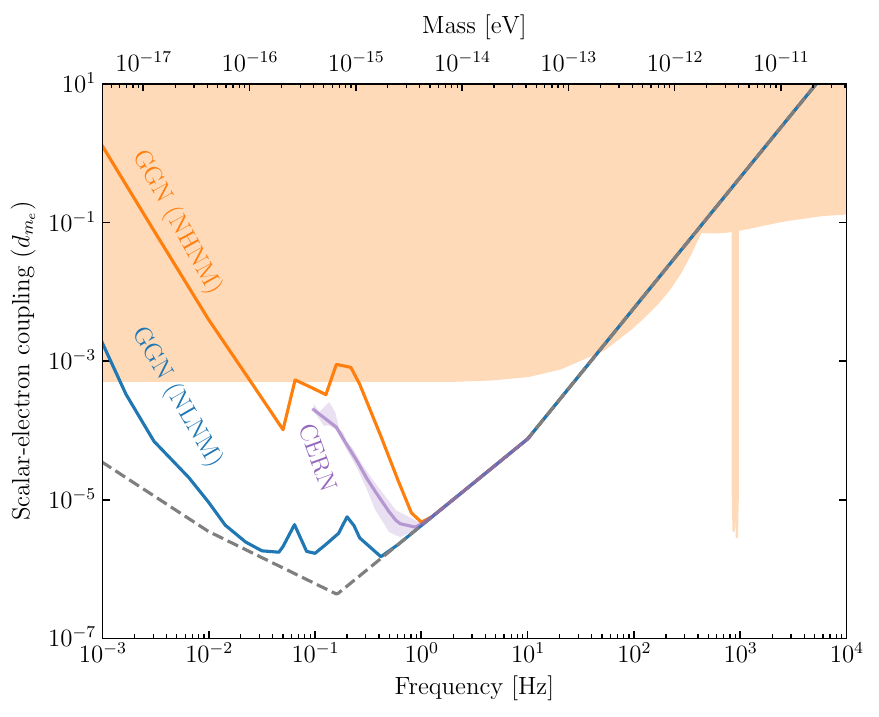}
 \includegraphics[width=0.50\textwidth]{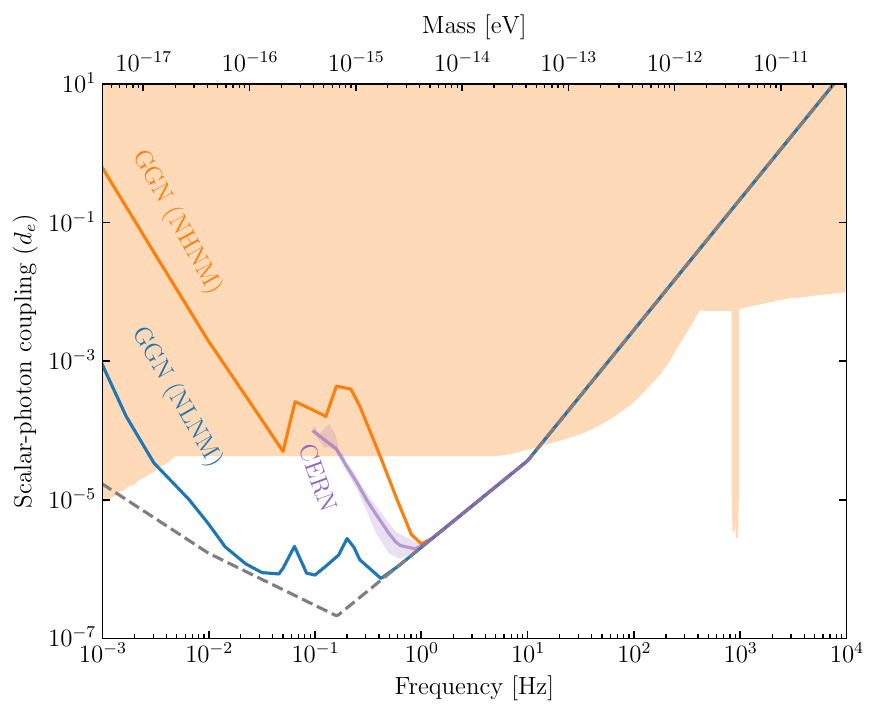}\\
 \includegraphics[width=0.48\textwidth]{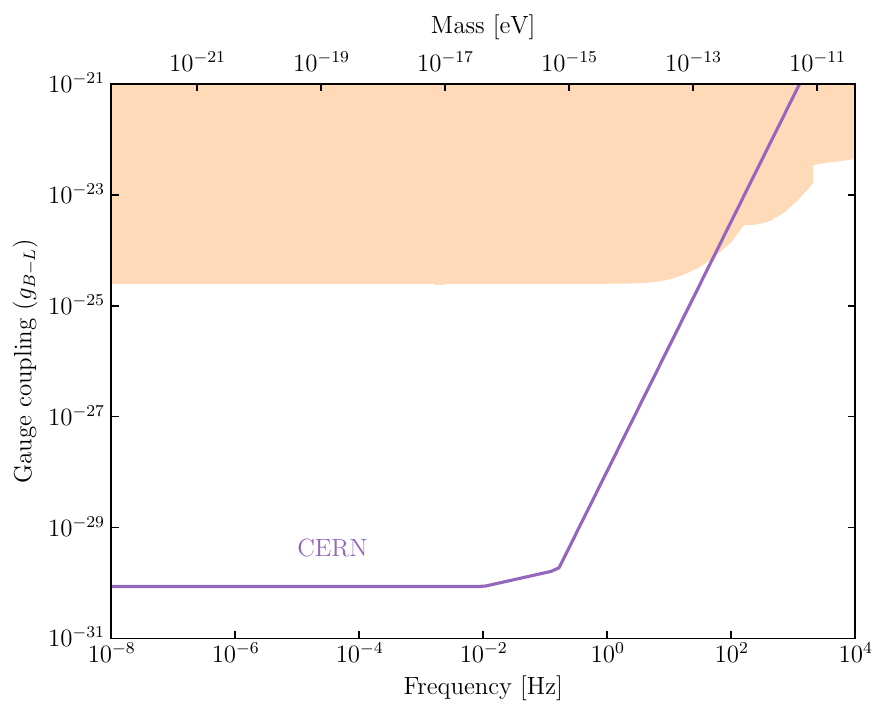}
 \includegraphics[width=0.50\textwidth]{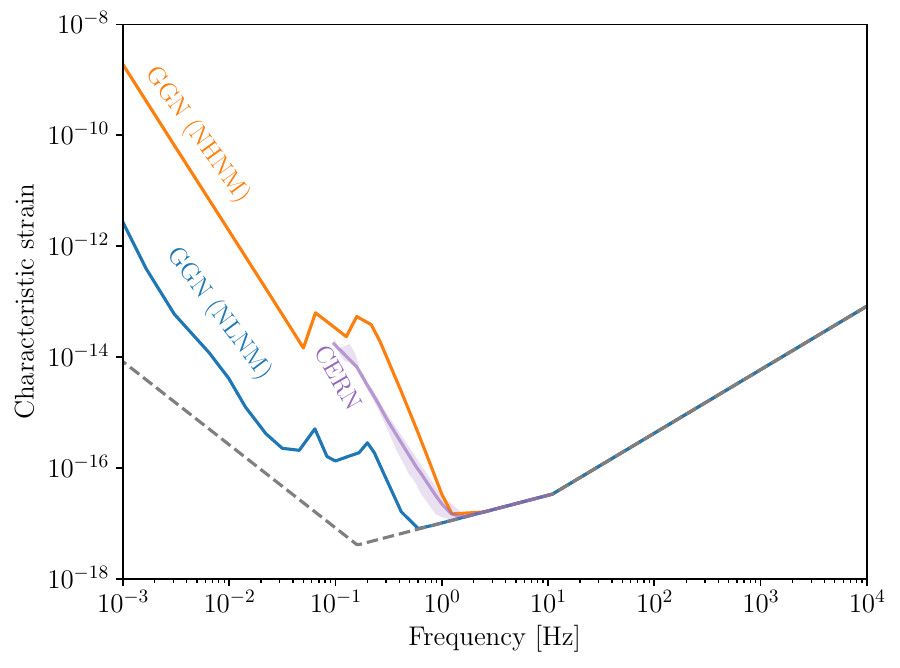}
 \caption{ \label{fig:AION100Sensitivities}Sensitivities of
 a 100-m Atom Interferometer to the couplings of scalar ULDM to the electron ({\it upper left panel}), to the photon ({\it upper right panel}), and to vector dark matter coupled to $B - L$ ({\it lower left panel}). The {\it lower right panel} shows the sensitivity to GW strain. The Atom Shot Noise (ASN) calculated from the indicative experimental parameters in Table~\ref{tab:AION100parameters} is shown by black dashed lines, the new high- (low-)noise model (NHNM) (NLNM) for gravity gradient noise (GGN) is indicated by solid orange (blue) lines. The GGN calculated on the basis of seismic measurements at the PX46 site is indicated by violet lines, with shading corresponding to the diurnal fluctuations in vertical surface motion reported in~\cite{Arduini:2851946}.
 %shown in Figs.~\ref{fig:RMS powerSpectral Density Min Max Case} and \ref{fig:noisemodels}.
 The surrounding rock is assumed to have properties similar to sandstone (molasse). Plots taken from~\cite{Arduini:2851946}.}
\end{figure*}
%%%%%%%%%%%%%%%%%%%%%%%%%%%%%

A coherent wave of ULDM may alter the excited atomic energy level, modifying the atomic phase and hence the interference pattern, which would also be  sensitive to the distortions of space-time caused by the passage of a GW. Figure~\ref{fig:AION100Sensitivities} illustrates the capabilities of a 100-m vertical AI detector assuming the indicative experimental parameters listed in Table~\ref{tab:AION100parameters}, and broadband operation. The upper panels illustrate the sensitivities of such an AI experiment to the couplings of scalar ULDM to Standard Model particles~\cite{Arvanitaki:2016fyj}, see~\cite{Badurina:2019hst,Badurina:2021lwr} and~\cite{Badurina:2022ngn}.
The lower left panel of Figure~\ref{fig:AION100Sensitivities} illustrates the potential sensitivity
to ultralight vector dark matter coupled to $B - L$, assuming two co-located interferometers using $^{88}$Sr and $^{87}$Sr isotopes, and 
the lower right panel of Figure~\ref{fig:AION100Sensitivities} shows the potential sensitivity to GW strain of a 100-m AI. The shaded regions in the ULDM panels are excluded by present measurements, including by MICROSCOPE~\cite{MICROSCOPE:2022doy}, torsion balance experiments~\cite{Hees:2018fpg}, LIGO and Virgo~\cite{LIGOScientific:2021ffg}, AURIGA~\cite{Branca:2016rez} and atomic clock experiments~\cite{Filzinger:2023zrs}.

{\it We emphasise that no other planned experiment would be as sensitive as a 100-m AI for probing the `mid-mass gap' for scalar ULDM searches, and that a 100-m AI would pioneer searches for GWs in an interesting frequency range inaccessible to laser interferometers.}

%In this case, in addition to the NHNM model for the GGN (solid orange line) we also display the New Low-Noise Model (solid blue line) as well as an estimate of the GGN based on seismic measurements in PX46. We see that the NLNM and PX46 calculations yield noise levels that are very close to the ASN over the frequency range displayed.

Additional science objectives for a 100~m AI experiment include searches for other forms of coupling between ULDM and SM particles~\cite{Badurina:2019hst,Beadle:2023flm,Blas:2024kps}, probing the Equivalence Principle and fifth forces~\cite{Biedermann:2014jya,Rosi:2017ieh}, constraining models of dark energy~\cite{Burrage:2014oza,Hamilton:2015zga,Sabulsky:2018jma} and the possibility of measuring the
gravitational Aharonov-Bohm effect, as pioneered in~\cite{Overstreet:2021hea}.

%These examples illustrate the potential scientific capabilities of a 100~m vertical AI experiment that could be located in an LHC access shaft. In the following Section of this report, we give an overview of such an experiment and the required technical infrastructure, and the following Section assesses the feasibility of installing such an experiment in the PX46 access shaft.

%~~\\
%~~\\
%~~\\

%% file: 4-FeasibilitySummary.tex
\section{Summary of the feasibility study}
%(S.~Calatroni)
\label{sec:Overview}

% \textit{1-2 pages: in particular, mention that PX46 can be separated from the LHC, allowing further works in total safety and in conjunction with LHC operation}
%\\

The previous conceptual feasibility study~\cite{Arduini:2851946} found no technical obstacle to siting a vertical AI at the LHC Point 4 inside the access shaft PX46, illustrated in Figs.~\ref{fig:PX46situation} and \ref{fig:artistic}. 
The reader is referred to the report for technical details, and we summarise here the main conclusions. The presence of the LHC infrastructure brings several advantages in terms of logistics for hosting an experiment, as well as some constraints that have to be managed for building and operating an AI experimental facility safely and reliably.

\begin{comment}
\begin{figure}
    \centering
    \includegraphics[width=0.49\linewidth]{Figures/Point4-extended_square.jpg} \includegraphics[width=0.49\linewidth]{Figures/Proposed design cropped.png}
    \caption{Schematic view of the underground areas around the LHC Point 4, and artistic view of the proposed implementation of a vertical AI in the PX46 shaft}
    \label{fig:point4}
\end{figure}
\end{comment}

As a pre-requirement, environmental studies were performed to ensure that the location and the existing infrastructure do not generate unwanted levels of background noise or perturbations on an AI experiment, in particular in terms of vibrations and seismic noise, and of electromagnetic noise and LHC-induced EMI. The level of all possible such perturbation sources has been found lower than what is required for an AI experiment, and compatible with simple mitigation measures when needed. Extensive details are reported in~\cite{Arduini:2851946}.

The main requirement that guided the feasibility study is that an AI experiment must be able to run at all times concurrently with the LHC, allowing personnel intervention for construction and commissioning of the experiment and for routine operation or maintenance tasks. This required a careful analysis of the hazards that might arise for the AI personnel and infrastructure from the LHC environment, while at the same time guaranteeing an adequate protection of the LHC itself from unwanted interferences.
We highlight here the key results of the thorough investigation performed with the contribution of several different CERN services and experts of the potential sources of hazard, and the practical mitigation measures that have been identified. The main points addressed were:
\begin{itemize}
    \item Radiation hazards and protective measures
    \item Fire safety and protective measures
    \item Helium release incidents
\end{itemize}

Radiation hazards of the AION experiment at LHC Point 4 have been preliminarily assessed in~\cite{Maietta:2020}. In the case of an accidental LHC beam loss in the vicinity of Point 4, the effective dose received at the bottom of the PX46 shaft would exceed the annual dose limit for an exposed person of 20 mSv, thus requiring the creation of a 80-cm thick concrete shielding wall~\cite{CERN-PBC-Notes-2022-003} in order to classify the PX46 shaft as a simple Supervised Radiation Area~\cite{EDMS-810149:2007} also during LHC operation. The shielding wall should have an opening of sufficient height and width to allow the passage of equipment for the LHC. The lower part of this opening would normally be closed by a movable door on rails, made of concrete blocks for ease of fabrication. The top part of the opening, expected to be used only upon dismantling or major refurbishments of the LHC, would be closed by concrete blocks stacked on a supporting beam. At the bottom of the shielding wall a set of access door to the LHC, interlocked and protected by a shielding chicane, would serve as emergency exit from the PX46 shaft. 

If a fire is detected in PX46 or in one of the  underground areas neighboring the RF installations (UX45, RUX45), personnel must evacuate using the elevator platform that is used to access the AI and reach the top of the shaft or, should evacuation to the surface not be possible, reaching the bottom of the shaft. The platform studied is electrically powered and has backup batteries. If these batteries fail, the platform has been designed to be able to perform a controlled mechanically braked descent to the bottom of the shaft in less than 2 minutes. AI operators should then evacuate through the access door at the bottom of PX46 using one of the existing LHC escape routes, either via TX46/UP46 to reach PM45 or via TX46/UX45 to reach PZ45, depending on the location of the fire and on the instructions received from the firefighters~\cite{LHC-0000006238:2013}. The 2-minute evacuation time as been assessed in~\cite{Arduini:2851946} and does not introduce a significant risk for the AI personnel.

Both the LHC magnet system and SRF accelerating cavity system are cryogenically cooled with liquid helium. In the event of an accidental helium release from the magnet cryostat, the helium would be guided through a set of pressure-resistant doors designed to cope with the helium flow of \SI{166}{m^3/sec} expected in a Maximum Credible Incident (MCI) during \SI{30}{sec}, and released through the PM45 shaft \cite{Gaignant:2021}, thereby not impacting the AI and its personnel. In case of accidental helium release from the SRF cavities, it would  flow from the cryostats in RUX45 to PX46 through UX45 and TU46. The total gas volume is however small compared to the UX45 cavern volume. This has been proven in practice during an incident where oxygen levels were monitored and found remaining within acceptable limits \cite{Hakulinen:2022}.

Technical services were also assessed in detail \cite{Arduini:2851946}. 

Existing Heating, Ventilation and Air Conditioning (HVAC) equipment for the LHC will not impact the AI, nor would the AI construction impact HVAC operation, provided the existing top closure of the PX46 shaft is maintained. Existing chilled water production in the neighbouring surface building SU4 does not have sufficient extra capacity for adding the AI as new user. A new station would have to be installed, which would allow the AI to operate independently from the LHC and its technical stops, but this could be done in parallel with the AI experiment construction. 

There is already enough electrical power available for an AI at Point 4, with a \SI{1.25}{MVA} transformer compared to a measured consumption of less than \SI{100}{kVA}. No major interventions are required for the works described in this report, except for the installation of the powering for the elevator, which would allow any interventions needed for the AI to be postponed to a later stage.

The above considerations led to a positive conclusion on the feasibility of an AI at LHC Point 4~\cite{Arduini:2851946}. Moreover, it  also emerged clearly that once the major works for making the PX46 shaft independent from the LHC (construction of the shielding wall, installation of an access control system, realization of the elevator platform) are completed all subsequent work for constructing and commissioning an AI could be carried out concurrently with the LHC operation. A working group was established at the beginning of 2025 with the mandate of defining the technical construction schedule for making PX46 independent from the LHC and to assess whether this would be feasible during LS3. An update of the design as well as preparing a more refined cost estimate was also within the scope of the working group activities. The conclusions of this study are documented in the present report.

%The results in case of an accidental HL-LHC beam loss at point 4 after implementation of the latest proposal for shielding in TX46 are shown in Fig.~\ref{fig:lhc_ir4_AION-100_RP_accidental}. The radiation levels are acceptable, even at the base of PX46. We note that, in addition, a radiation monitor IG5-H20 (CMPU-W) including an alarm unit (CAU) and uninterruptible power supply (CUPS-W) would be needed to monitor the radiation hazard in the PX46 shaft during LHC operation, for a cost estimate of 15 kCHF including installation activities, local cabling and commissioning. A formal request for instrumentation (RFI) would be submitted to HSE/RP-IL in due time, ensuring beforehand the availability of sockets for the IT and EN/EL networks.

%\begin{figure}
%    \centering
%    \makebox[\textwidth][c]{
%        \includegraphics[width=0.8\textwidth]{Figures/RP/lhc_ir4_AION-100_RP_accidental_Ym600-m400_lines.pdf}
%    }
%    \makebox[\textwidth][c]{
%        \includegraphics[width=1.2\textwidth]{Figures/RP/lhc_ir4_AION-100_RP_accidental_X1900-2100_lines.pdf}
%    }
%    \caption{Ambient Dose Equivalent in TX46/PX46 from an accidental HL-LHC beam loss at Point~4 after implementation of the latest shielding design in TX46. {\it Upper panel}: a transversal cut centred at the emergency door level. {\it Lower panel}: a longitudinal cut centred in the middle of TX46. Contour lines in blue correspond to the yearly limit for Supervised Radiation Areas.}
%    \label{fig:lhc_ir4_AION-100_RP_accidental}
%\end{figure}

%% file: 5-CivilEngineering.tex
\section{Civil engineering - shielding wall}
%(T.A. Bud)
\label{sec:CE}
\newcolumntype{L}[1]{>{\raggedright\arraybackslash}p{#1}}
%2-3 pages: description of the needed CE infrastructure to be built and potential timeline (design studies, external contractor, etc.). Refined cost estimate.

Minor civil engineering works will be required to house an AI experiment in the proposed location. 

\begin{figure}[h!]
     \centering
     \includegraphics[width=.45\textwidth]{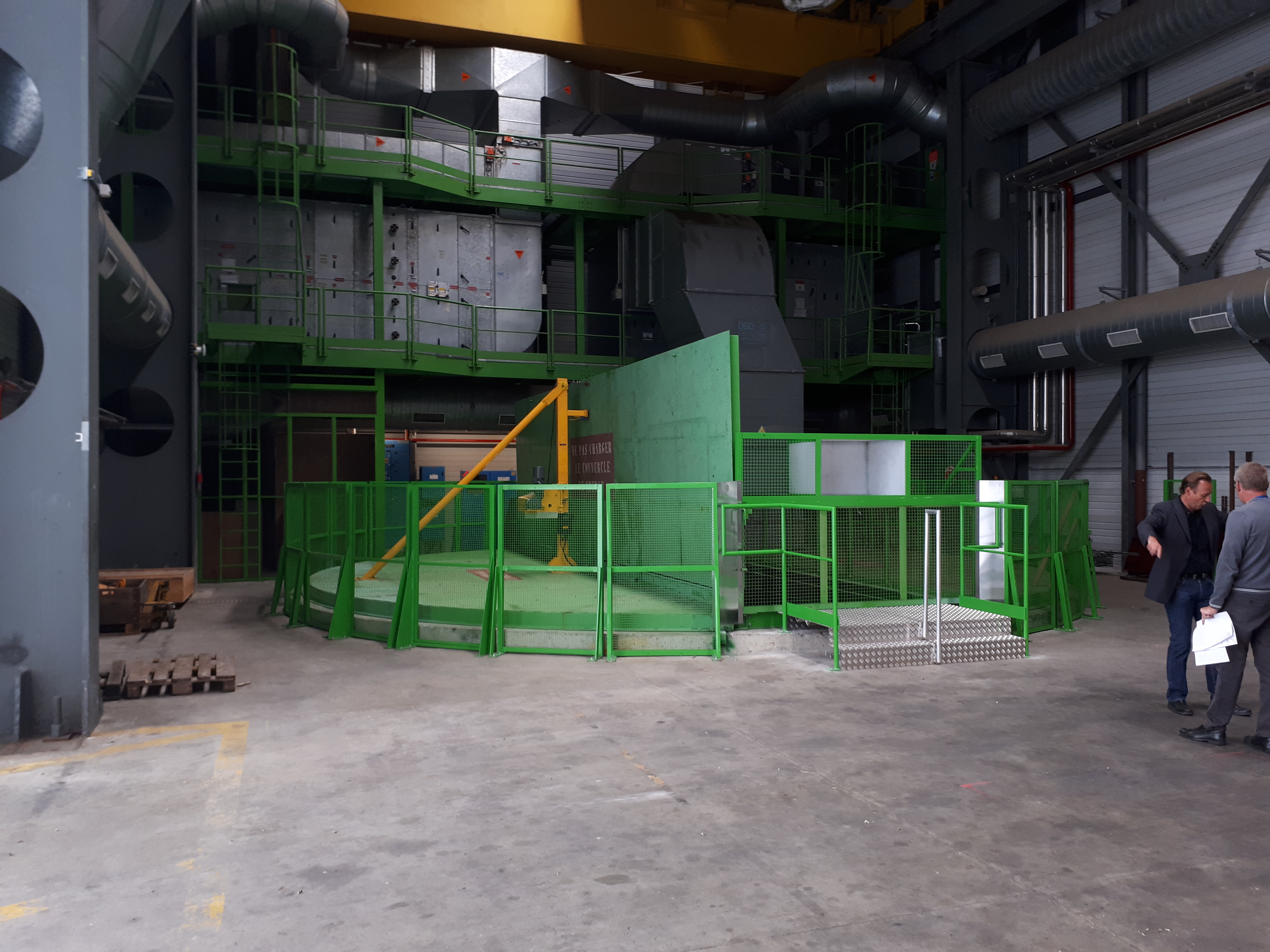}
     \quad
     \includegraphics[width=.45\textwidth]{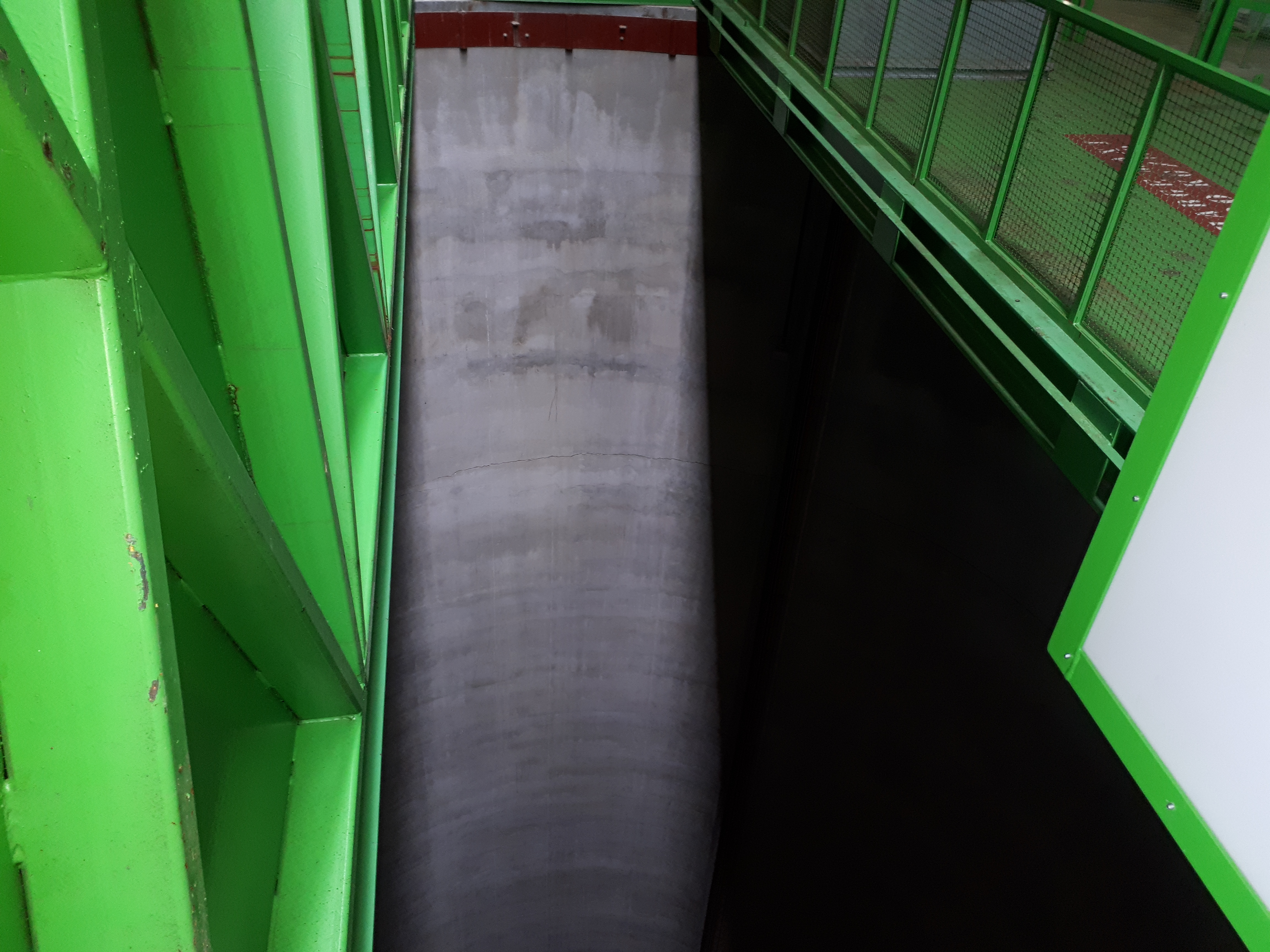}
     \caption{\label{fig:CE:steel cover} View from the building SX4 showing the existing steel cover at the top of the PX46 shaft.}
 \end{figure}

Fig.~\ref{fig:artistic} shows the existing infrastructure at Point 4 of the LHC tunnel with the proposed AI located in the PX46 shaft.
Currently the top of the shaft is closed with a steel cover that can be opened in the middle, with a hoist to allow transport into and out of the shaft, see Fig.~\ref{fig:CE:steel cover}. This will require some modification of the design to allow the installation of the experiment with the proposed elevator platform and regular access from the top. A new enclosed access area will be created for this purpose with a door controlled by a badge reader connected to the LHC Access Control System~(see Sec.~\ref{sec:Access}).

\begin{comment}
\begin{figure}[h!]
    \centering
    \includegraphics[width=0.8\textwidth]{Figures/Feasibility_Study/7-CE/Proposed design cropped.png}
    \caption{\label{fig:CE:proposed} 3D model of the existing infrastructure at Point 4 of LHC with the proposed AI located in the PX46 shaft and the shielding required in TX46.}
\end{figure}
\end{comment}

The main conclusion of the Radiation Protection studies described in the conceptual feasibility study~\cite{Arduini:2851946} was that installation of a shielding wall will be required for regular access, including during LHC beam operations, and to be able to use the full depth of the shaft. 
%Initially two different solutions were proposed, one having a shielding wall in the TX46 tunnel and the other having a shielding slab at the bottom of the shaft. Following analysis of both proposed solutions, it was determined that the only practically feasible option was the former, i.e., shielding in TX46. 

In the TX46 tunnel a 0.8~m thick cast-in-place concrete wall is proposed, with removable shielding blocks covering approximately 28~m$^2$ that can be dismantled during long shutdowns, freeing the area for transportation and handling. Of this 28~m$^2$, 16~m$^2$ will be designed to move frequently (e.g., during end-of-year shutdowns or during LHC technical stops as required) and therefore installed on a motorized frame (see Sec.~\ref{sec:Platform}). The remaining 12~m$^2$ of shielding blocks are expected to be moved only once, during disassembly of the LHC, with mobile lifting equipment. 

\begin{figure}
    \centering
    \makebox[\textwidth][c]{
        \includegraphics[width=0.8\textwidth]{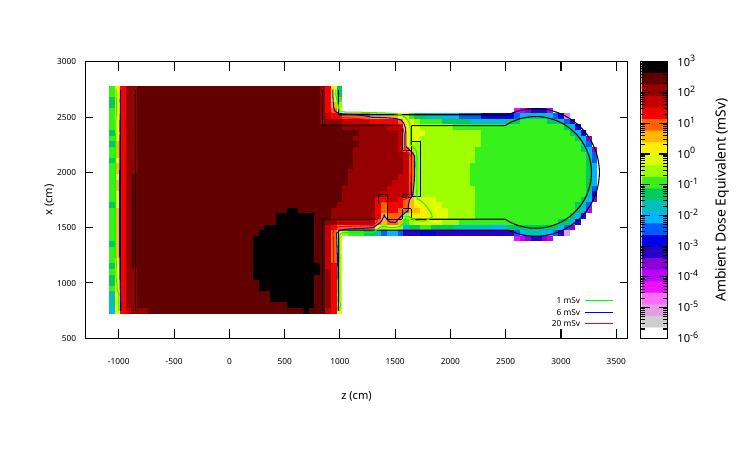}
    }
    \makebox[\textwidth][c]{
        \includegraphics[width=1.2\textwidth]{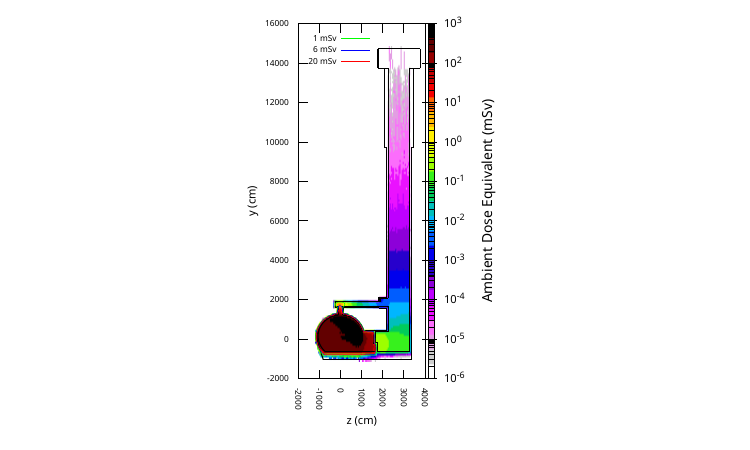}
    }
    \caption{Ambient Dose Equivalent in TX46/PX46 from an accidental HL-LHC beam loss at Point~4 after implementation of the latest shielding design in TX46. {\it Upper panel}: a transversal cut centred at the emergency door level. {\it Lower panel}: a longitudinal cut centred in the middle of TX46. Contour lines in blue correspond to the yearly limit for Supervised Radiation Areas.}
    \label{fig:lhc_ir4_AION-100_RP_accidental}
\end{figure}

\begin{figure}[h!]
    \centering
    \includegraphics[width=0.8\textwidth]{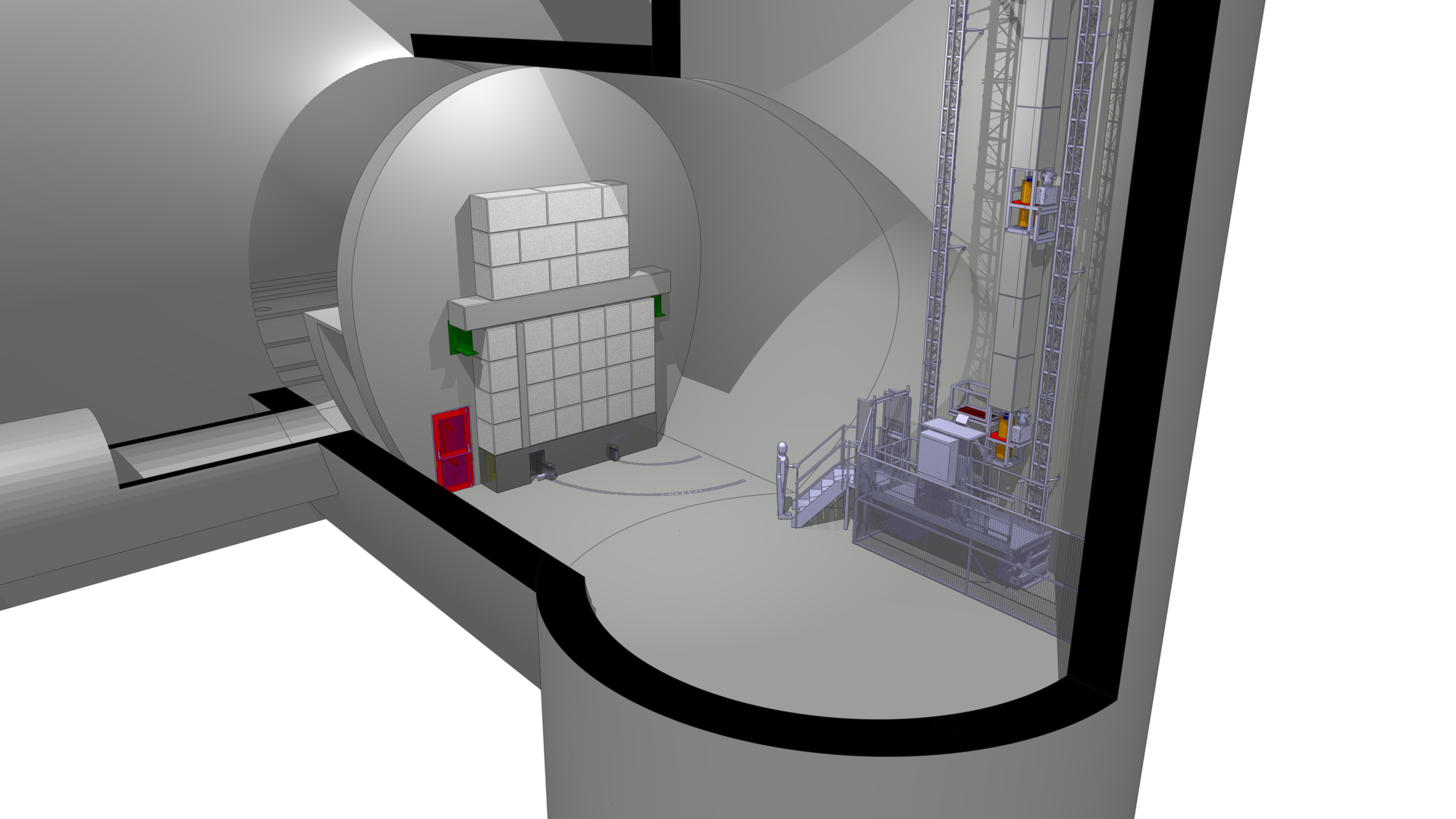}
    \includegraphics[width=0.8\textwidth]{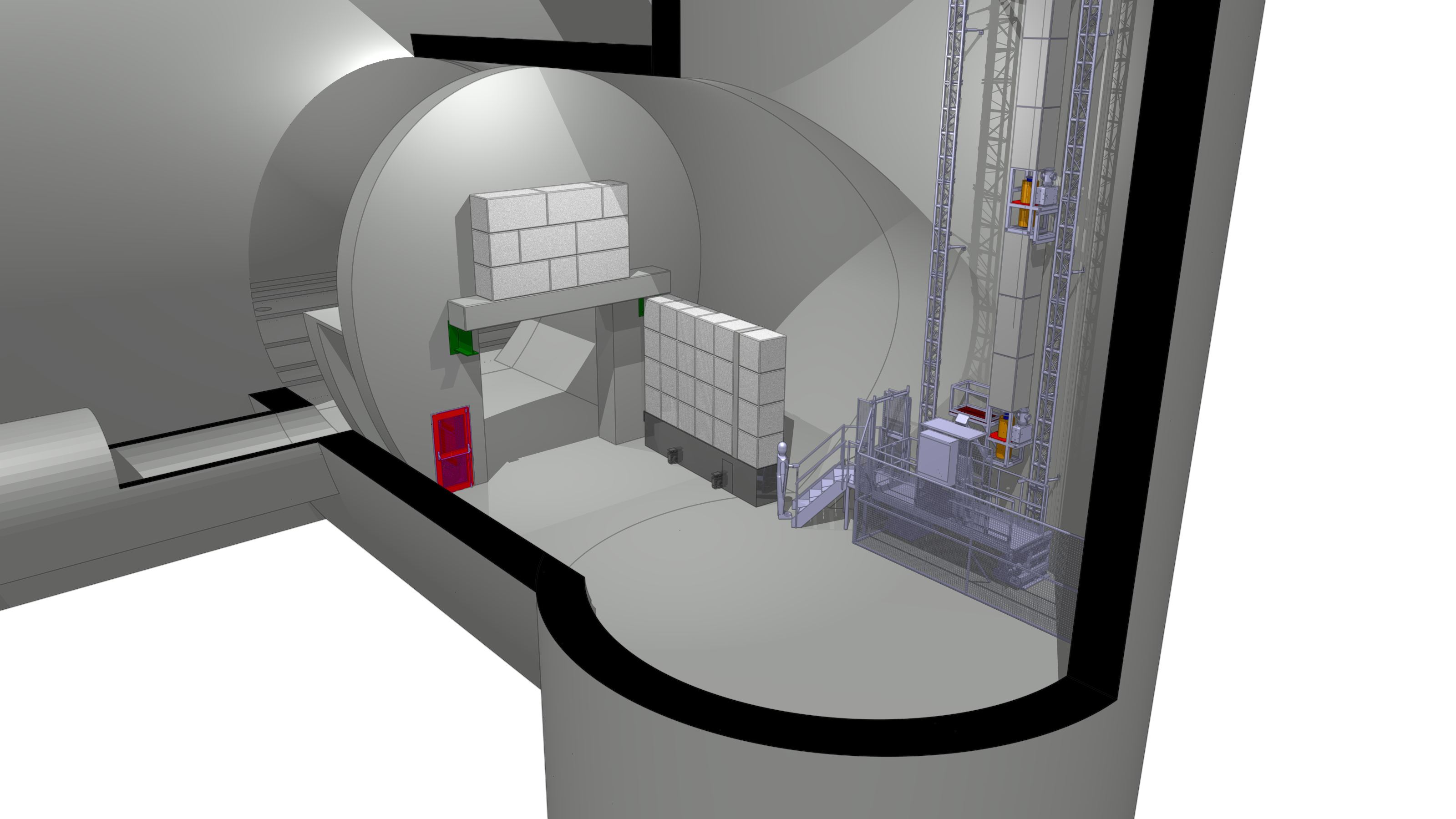}
    \caption{\label{fig:CE:chicane}Shielding wall with the proposed access doors in closed ({\it Upper panel}) and open ({\it Lower panel}) configuration. The bottom landing of the elevator platform is also visible.}
\end{figure}

The results in case of an accidental HL-LHC beam loss at point 4 after implementation of the latest proposal for shielding in TX46 are shown in Fig.~\ref{fig:lhc_ir4_AION-100_RP_accidental}. The radiation levels are acceptable, even at the base of PX46. We note that, in addition, a radiation monitor IG5-H20 (CMPU-W) including an alarm unit (CAU) and uninterruptible power supply (CUPS-W) would be needed to monitor the radiation hazard in the PX46 shaft during LHC operation. 
%for a cost estimate of 15 kCHF including installation activities, local cabling and commissioning. 
A formal request for instrumentation (RFI) would be submitted to HSE/RP-IL in due time, ensuring beforehand the availability of sockets for the IT and EN/EL networks.

An escape route from the bottom of the shaft is envisaged via the existing UP46 side tunnel. In order to provide a safe evacuation path through the shielding wall, it is proposed to install two new access doors in the shielding wall in TX46: an end-of-zone door and a second ventilation door with an additional 0.8~m thick chicane wall, as shown in Figs.~\ref{fig:CE:chicane} and \ref{fig:AC:lasscontacts}. 

\begin{figure}[h!]
    \centering
    \includegraphics[width=0.8\linewidth]{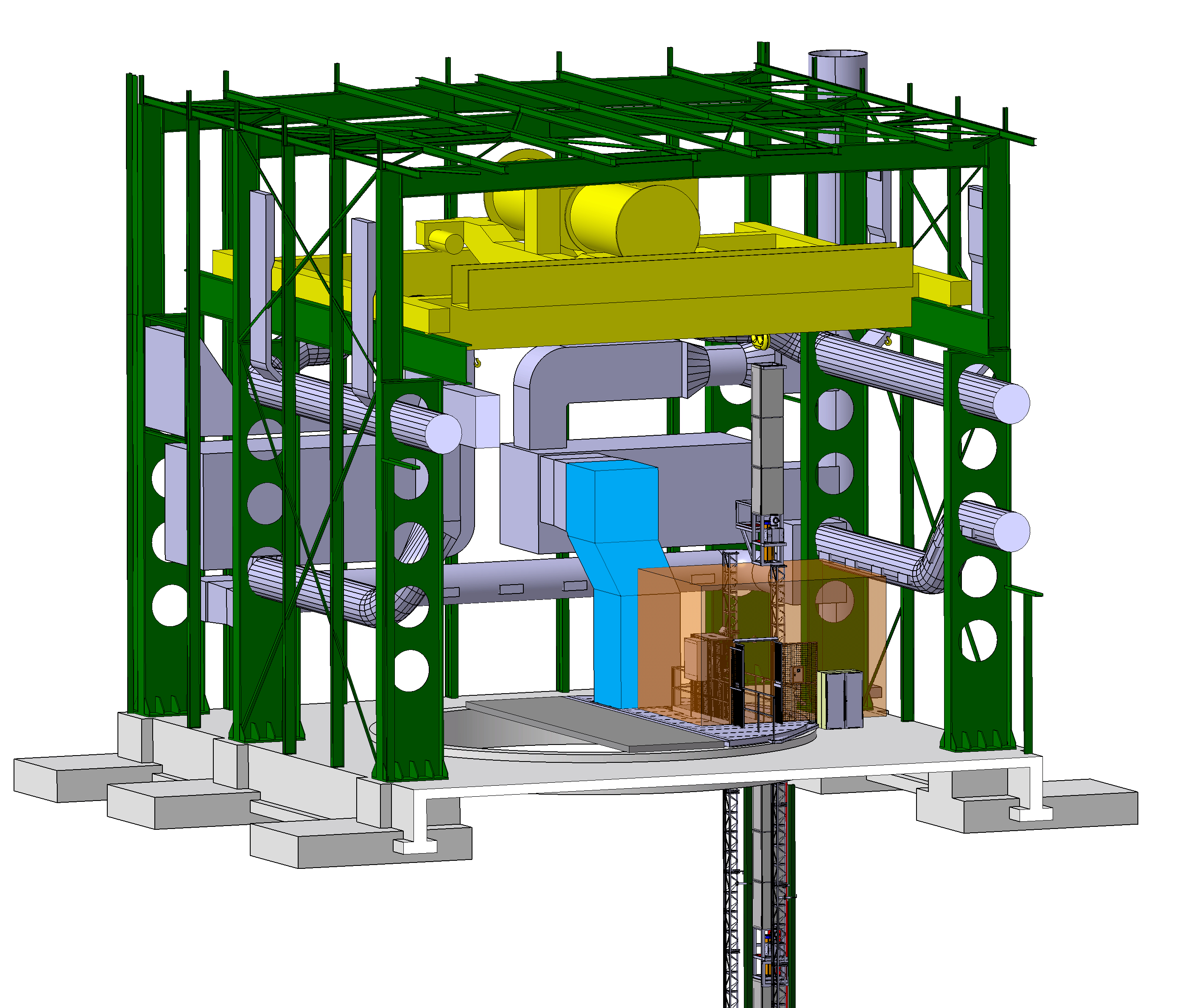}
    \caption{Ventilation room in the existing surface building SX4. The room can easily be enlarged in a second phase to host the laser laboratory.}
    \label{fig:SX4 vent building}
\end{figure}

On the surface, an approximately 20~m$^2$ ventilation room will be built in the existing building SX4 on the top of the shaft to ensure air tightness around the elevator platform to maintain the slight overpressure inside the shaft guaranteeing unchanged ventilation management of the underground areas. The space is available and no technical issues are expected, with two cascading doors to minimize airflow when accessing the room. For the purposes of cost estimation at this stage of the study, the room has been assumed to be a steel framed building of dimensions 4m x 5m x 3.5m. The room can be seen in orange in Fig.~\ref{fig:SX4 vent building}, and it is foreseen that it could be enlarged in a second stage to host the laser laboratory for the AI.

A preliminary cost estimate for the civil engineering works was calculated in 2021 using the chicane shielding design as the basis. This has since been updated and inflation figures have been used to adjust the cost estimate to 2025 numbers. The updated cost estimate is provided in Sec.~\ref{sec:Cost}.

Alongside this cost estimate, a timeline estimate for the civil works has been developed. The timeline, also provided in Sec.~\ref{sec:Cost}, gives an estimate for the works involved in the procurement and construction of the shielding wall in TX46. 
%Due to the uncertainties around the mechanism for the movable blocks and the outcome of future site investigation works, it should be highlighted that this estimate is preliminary and may change in the future. 

%% file: 6-AccessControl.tex
\section{Access control and safety systems}
%(T. Hakulinen)
\label{sec:Access}

%\it 2-3 pages: technical description of the access control system, doors and interlocks needed for safe experiment and machine operation, including egress in emergency situations. Mention fire and ODH detection. Prepare potential timeline (design studies, external contractor, etc.) and refined cost estimate.

%\begin{figure}[htbp]
%     \centering
%     \includegraphics[width=.4\textwidth]{Figures/Feasibility_Study/5 Infrastructure and safety/badgereader.jpeg}
%     \quad
%     \includegraphics[width=.4\textwidth]{Figures/Feasibility_Study/5 Infrastructure and safety/padmad.jpeg}
%     \caption{\label{fig:AC:badgepadmad} LACS badge reader (left), PAD and MAD (right)}
%\end{figure}

\subsection{Access safety}

The required modifications to the access control system fall into two distinct categories: access control to the shaft PX46, which will normally be classified as a supervised radiation area, and access control to the surface building SX4 and the laser room adjacent to the shaft. For the purposes of this implementation study, only the modifications that need to be carried out during the LS3 to facilitate the subsequent installation and operation of the experiment will be considered. It will be possible to descend from the top of the shaft  to the bottom via the planned elevator, therefore all the sectorization changes in the underground areas UX45, TX46 and PX46 will need to be finalized beforehand.

\begin{figure}[htbp]
     \centering
     \includegraphics[width=0.8\textwidth]{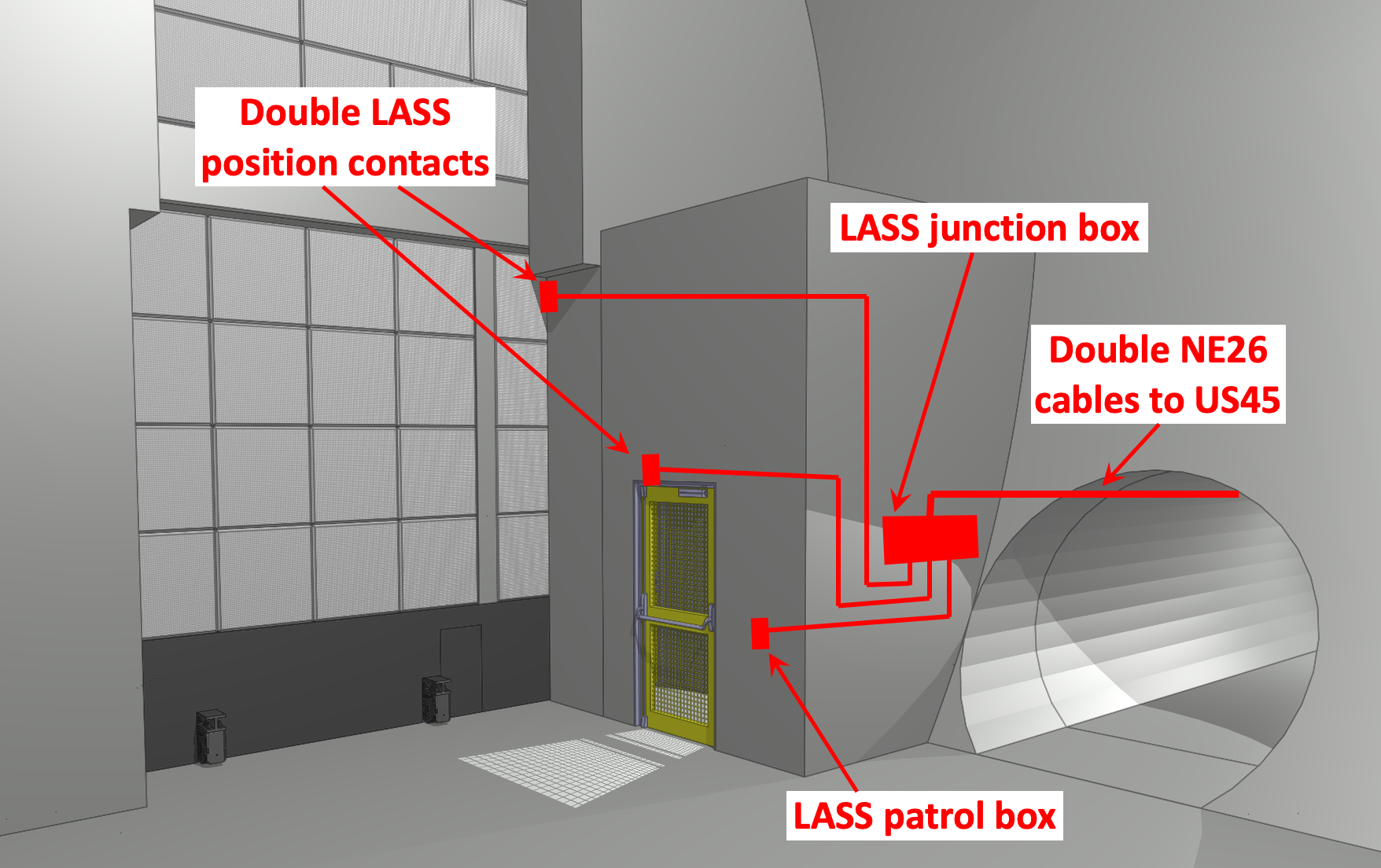}
     \caption{\label{fig:AC:lasscontacts} LASS instrumentation of access doors.}
\end{figure}

\begin{figure}[h!]
     \centering
     \includegraphics[width=.4\textwidth]{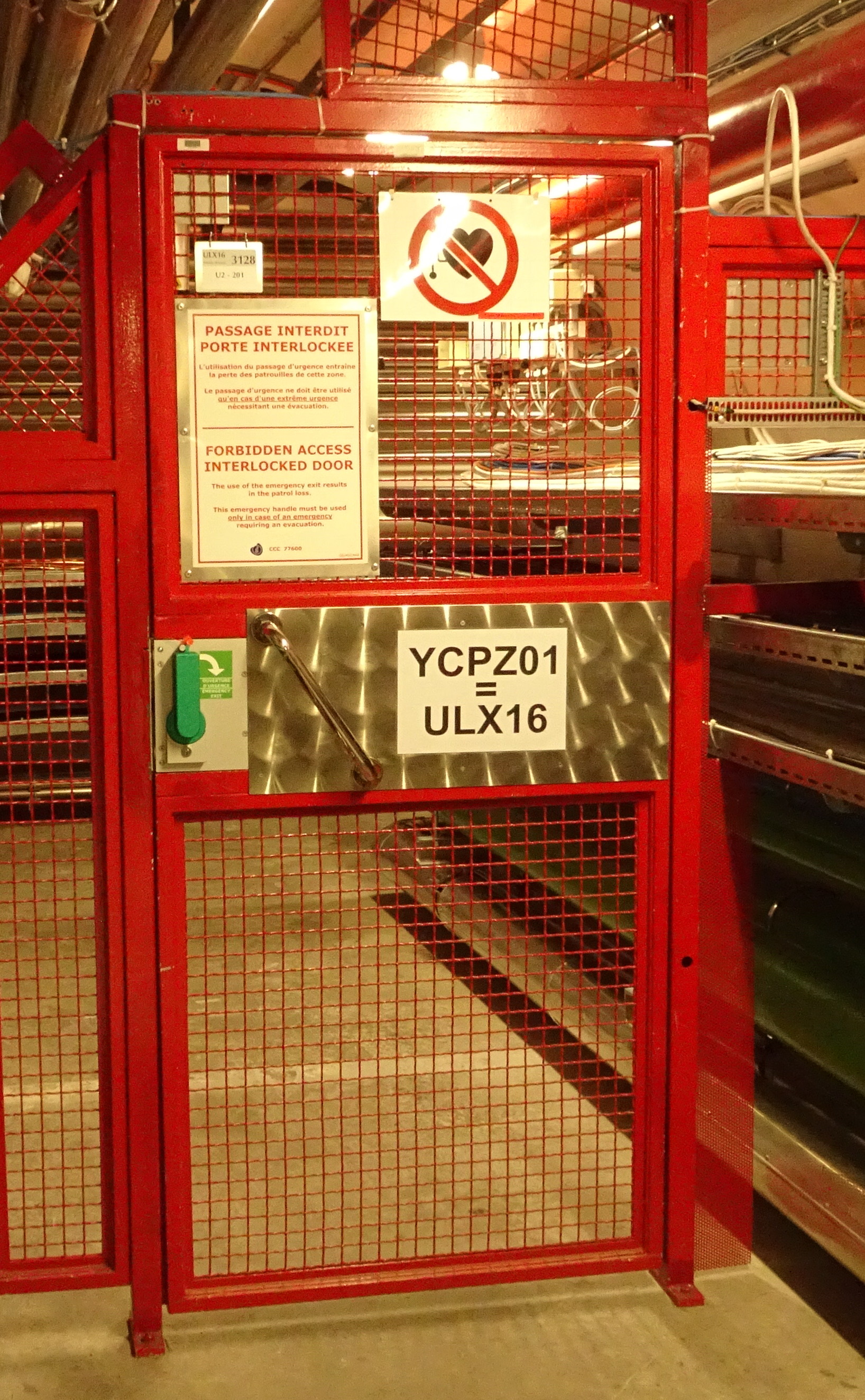}
     \quad
     \includegraphics[width=.4\textwidth]{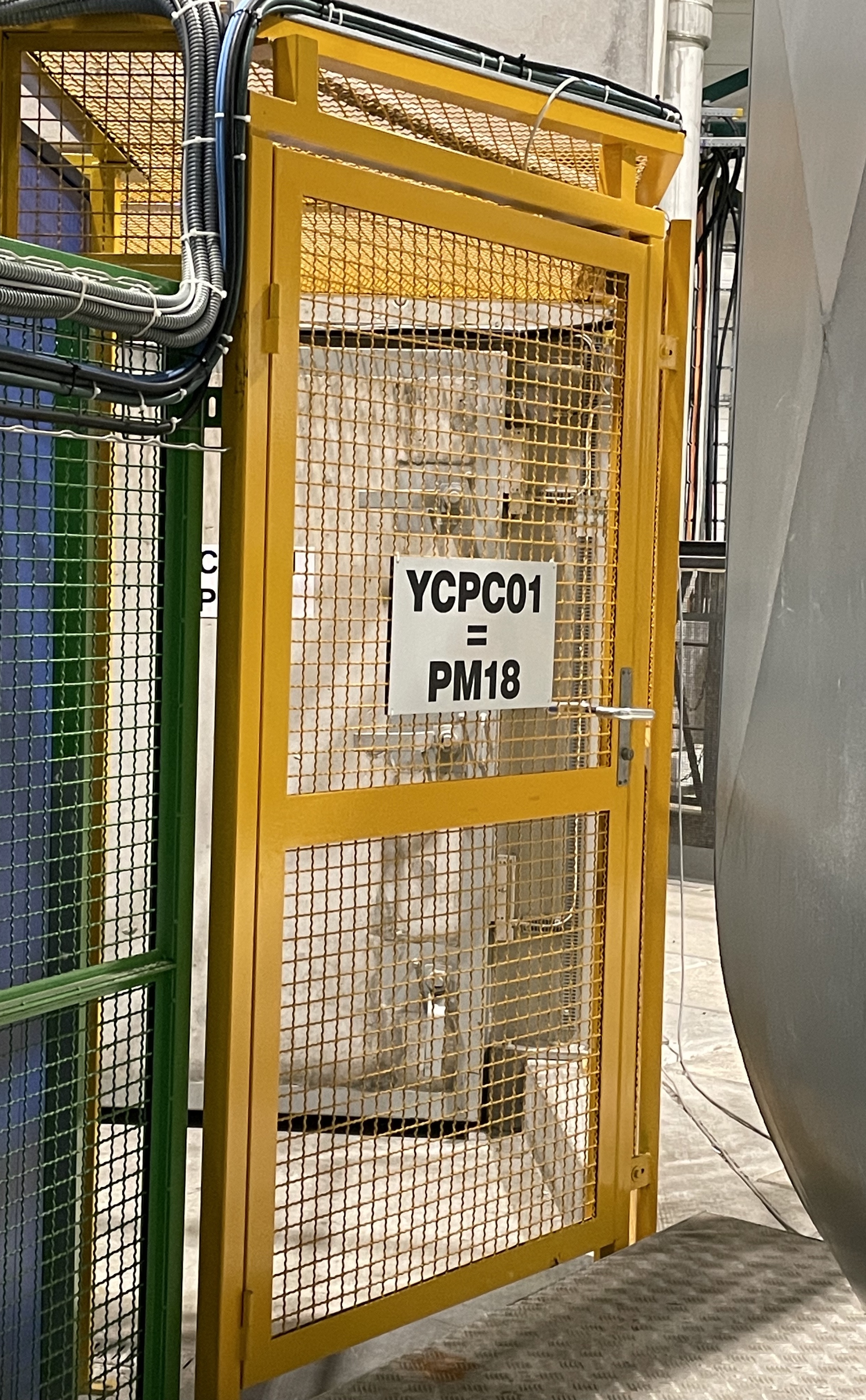}
     \caption{\label{fig:AC:eozventdoors} LASS end-of-zone door (left), and an example of a door for the second barrier (right).}
\end{figure}

The LHC Access Safety System (LASS) will be modified to accomplish the changes to the sectorization of the UX45 area.
Currently the entire PX46 shaft belongs to the interlocked LASS zone PZ45 together with the UX45 cavern. During an LHC run the top of the shaft is covered to prevent access and to contain any prompt radiation in case of an accidental beam loss close to UX45.
The new end of the PZ45 zone will be at the TX46 connecting gallery, where sufficient radiological shielding will be installed as per the analysis by the Radiation Protection team presented in the Conceptual Feasibility Study~\cite{Arduini:2851946} and refined in Sec.~\ref{sec:CE}.
As discussed there, two new access doors are to be installed successively in the passage through the shielding in TX46, an end-of-zone door and a second barrier door.
This arrangement follows the general principle of the LASS that any access from a non-interlocked area to an area where a radiological risk is present must pass through a minimum of two interlocked doors with separate safety contacts and cabling.
The end-of-zone door is a standard LASS door: red in colour, gridded, with double position contacts (connected to both LASS safety chains, PLC- and relay-based), and including an emergency opening handle on both sides (see Fig.~\ref{fig:AC:lasscontacts}).
The second barrier door is also gridded, like the end-of-zone door, with regular opening handles and LASS double position contacts (see Fig.~\ref{fig:AC:eozventdoors}).
If either of these doors is opened, or even if the end-of-zone door emergency handle is turned, the LHC beam will be dumped and the system put into a safe state.

As the most recent layout described in Sec.~\ref{sec:CE} calls for a fully movable shielding wall, similar to the large motorized shielding walls of the LHC experiments, LASS safety contacts are to be installed also on this device. However, as shielding walls are not considered as emergency exit paths, but rather condemned (with motors de-energized) during LHC operation, installation of a second barrier is not required.

Installation of an end-of-zone door in TX46 means that PX46 shaft can be classified as a non-interlocked area, no longer supervised by the LASS and allowing access from the top via a simplified procedure by the LHC Access Control System (LACS), to be implemented following the requirements of the experiment. 
This status is similar to other non-interlocked experiment areas at PX15 (ATLAS), PM54 (CMS), PZ85 (LHCb), as well as HL-LHC PM17 and PM57.
As described in Sec.~\ref{sec:CE}, TX46 will act as an evacuation route via the UX45 cavern if the access to the top of the shaft is unavailable or dangerous.

\subsection{Alarm systems}

\begin{figure}[h!]
     \centering
     \includegraphics[width=.45\textwidth]{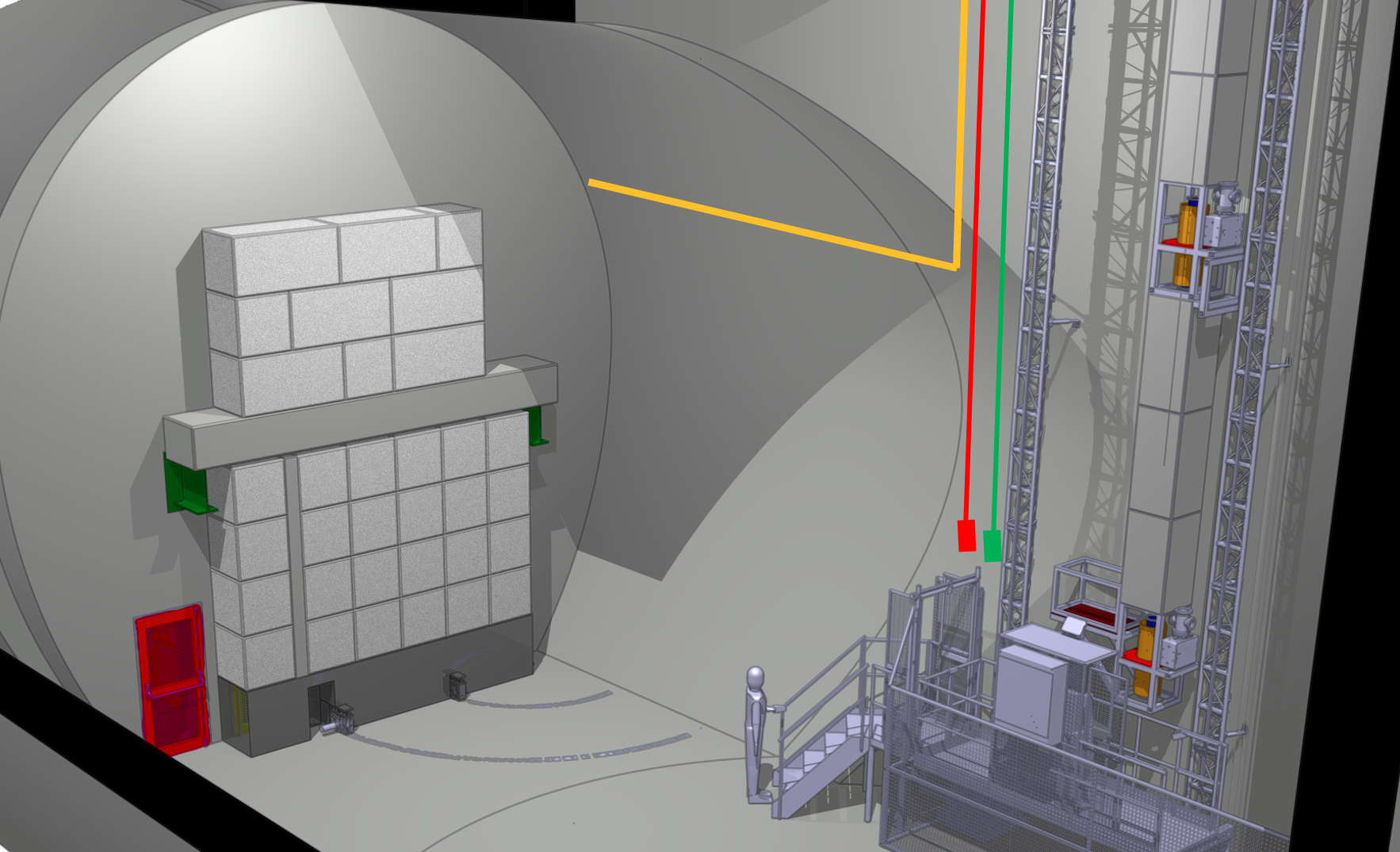}
     \quad
     \includegraphics[width=.45\textwidth]{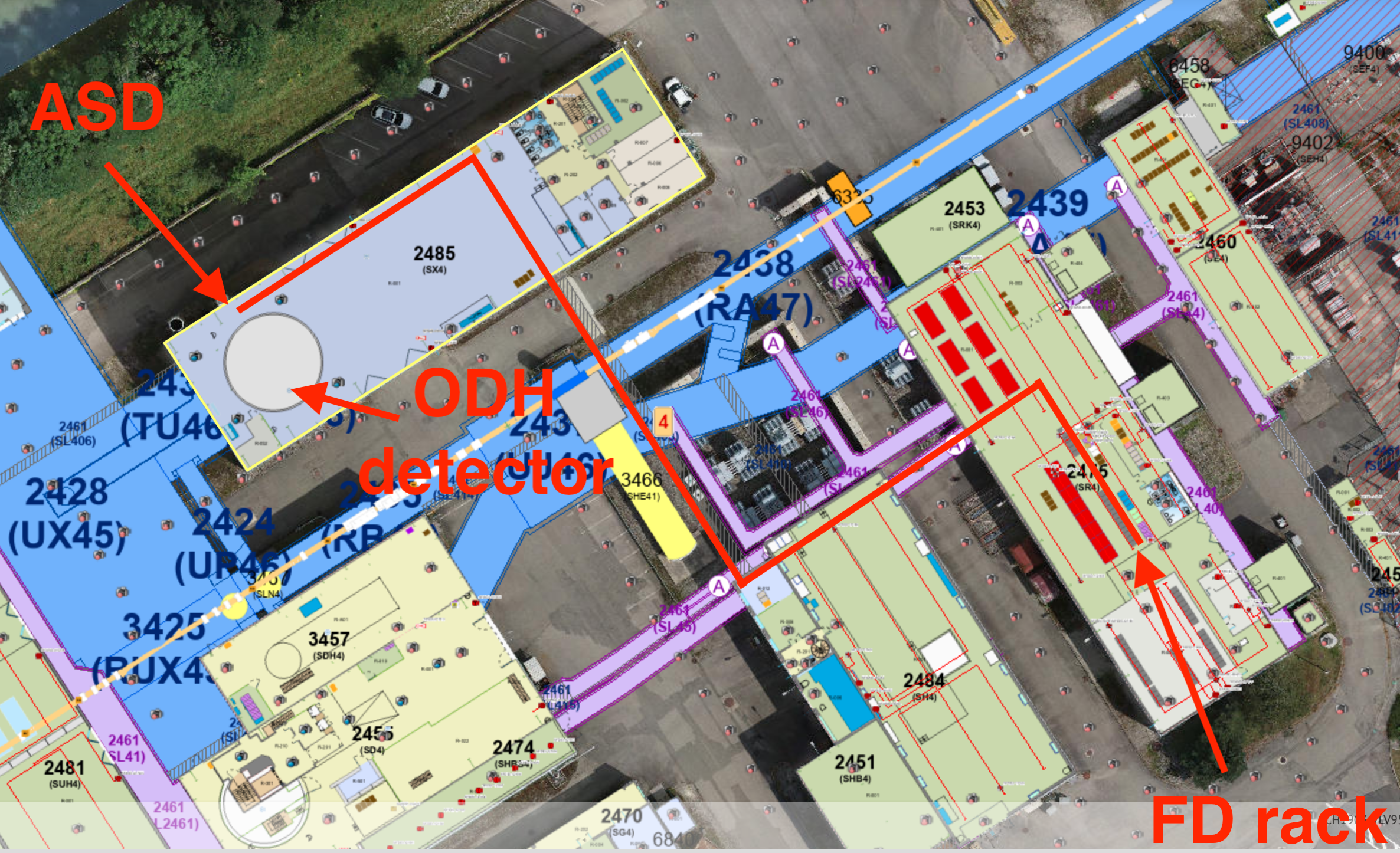}
     \caption{\label{fig:AC:firedet} Layouts of the fire detection aspiration tube (orange), red emergency telephone (red), and manual call point (green) at the bottom of the PX46 shaft (left panel), and cabling from the fire detection rack to the air aspiration device (ASD) (right panel).}
\end{figure}

\begin{figure}[h!]
     \centering
     \includegraphics[width=0.8\textwidth]{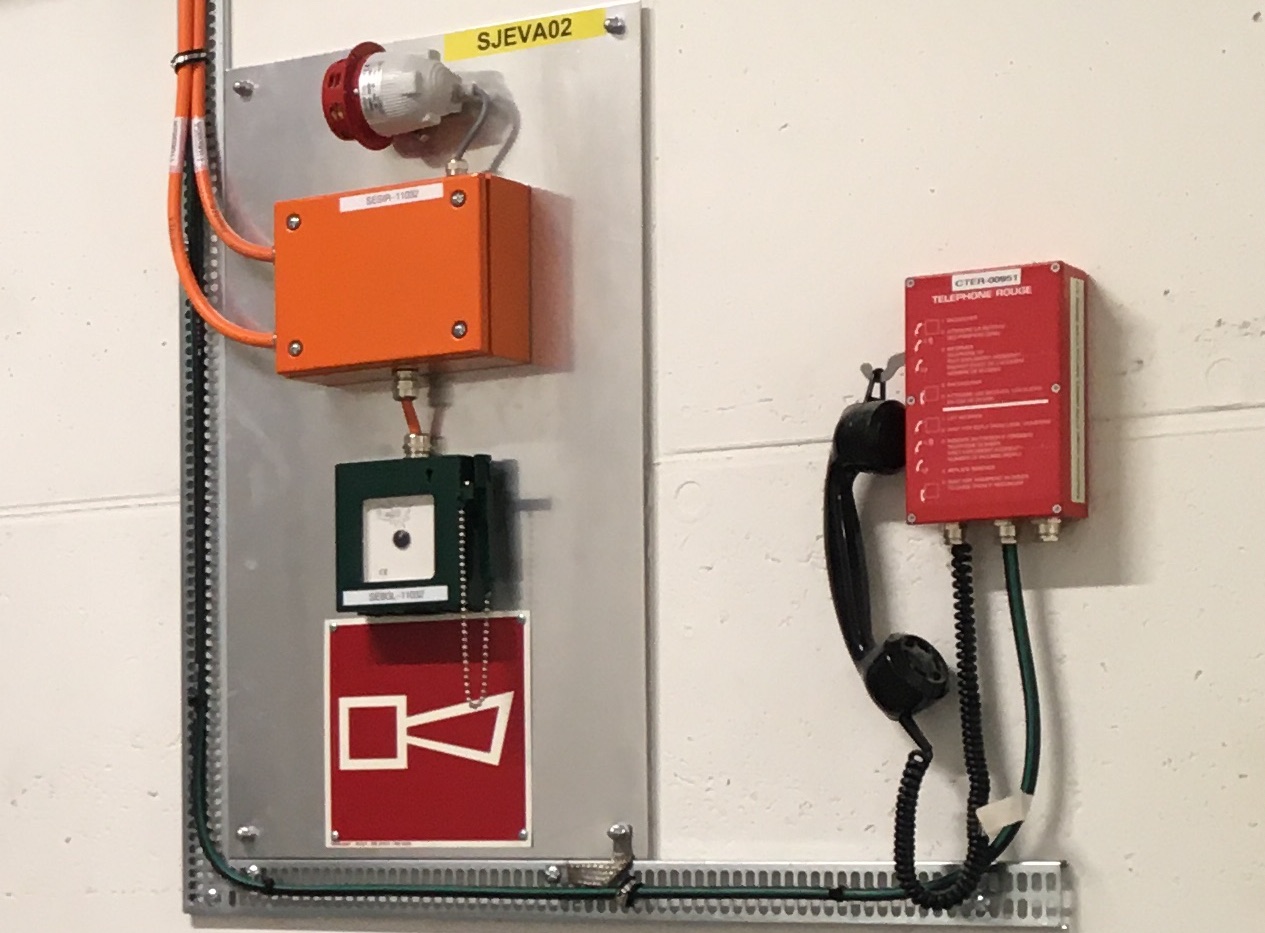}
     \caption{\label{fig:AC:evacandrt} Manual evacuation call point (``break-the-glass''), siren, and red emergency telephone.}
%     \centering
%     \includegraphics[width=.4\textwidth]{Figures/breaktheglass.png}
%     \quad
%     \includegraphics[width=.4\textwidth]{Figures/redtelephone.png}
%     \caption{\label{fig:AC:evacandrt} Manual evacuation call point (left panel) and Red Telephone (right panel).}
\end{figure}

Alarm systems for fire and oxygen deficiency hazards and evacuation as well as emergency communication will also be necessary. For the installation phase, basic services are to be installed during LS3, which can then be extended to cover any possible risks of the experimental phase. The alarm systems in question are:

\begin{itemize}

\item {\bf Fire detection:}
Smoke detection by aspiration tubes is currently the most effective method of early detection of fire. It is proposed to install a detection unit at the top of the shaft, from which detection tubes can be installed reaching the bottom of the shaft (see the left panel of Fig.~\ref{fig:AC:firedet}). Cabling for the fire detection system is pulled from the main Fire Detection System rack of Point 4 in building SR4 (see the right panel of Fig.~\ref{fig:AC:firedet}).

\item {\bf ODH detection:} 
Based on previous experience it is assessed that the probability of a large He-release (the most credible incident, MCI) in PX46 is small, as demonstrated by earlier real He-releases in the RF area, where the helium gas has been largely contained in the LHC tunnel and driven away from UX45 towards other LHC sites, due to the effect of the regular ventilation. Therefore, it is estimated at this time that the existing ODH detection already present at the top of the PX46 shaft and the SX4 building will be sufficient.

\item {\bf Evacuation:}
Any hazard detection will need to be communicated to the personnel present on site by the evacuation system. At this time a simple siren is foreseen to be installed at the top of the shaft, which will give a general evacuation signal in case of a fire or ODH detection. It will be the responsibility of the personnel to evaluate the situation and to decide the best evacuation option based on the location of the hazard, either via the elevator or through UX45. In addition, two manual call points, i.e., ``break-the-glass devices'' (see Fig.~\ref{fig:AC:evacandrt}), are to be installed in the shaft, one at the top and one at the bottom. Activating the call point by pushing the button will sound the evacuation alarm and alert the CERN Fire Brigade just as a fire or an ODH alarm would.

\item {\bf Emergency communication:}
A standard emergency communication device used in the LHC, a Red Telephone (see  Fig.~\ref{fig:AC:evacandrt}), is to be installed at the bottom of the shaft. Lifting the receiver will dispatch a Level 3 alarm to CERN Fire Brigade and open direct voice communication with CERN Safety Control Room (SCR).

\end{itemize}

The exact detection and alarm needs for the final experiment will be defined by a risk analysis by the competent safety officers.

\subsection{Budget and planning}

The currently foreseen budget for the modifications to be carried out during LS3 is presented in Sec.~\ref{sec:Cost}.

Planning of the works during LS3 is almost entirely tied to the advancement of the civil engineering and elevator platform works. Instrumenting the doors and the shielding wall obviously requires those elements to be fully commissioned, and it is not feasible to install fire detection in the shaft before the lifting platform is in place. However, there is certain latitude in the cabling works, in particular of the LASS trunk cabling and alarm system cabling from their respective control racks.

\vspace{0.5cm}

%Edit the file 5-4-3-AccessControl.tex to insert your contribution

%Figures should be stored in the folder Figures/5-4-3-AccessControl

%Example of citation~\cite{Witten:1998qj}. 
%Whenever possible, please use the Inspires texkeys. This makes it easier to avoid duplicate ciations in the bibliography.

%% file: 7-LiftingPlatform.tex
\section{Handling engineering equipment}
%(D. Lafarge)
\label{sec:Platform}

\subsection{Elevator platform and handling equipment}

The handling equipment required in order to implement a ${\cal O}(100)$~m AI experiment in the PX46 shaft consists of:
\begin{itemize}
    \item Specific lifting equipment to handle pieces of experiment very close to the walls of the PX46 shaft;
    \item An elevator platform providing access to the experiment at all levels;
	\item Mobile shielding to allow the separation from the UX45 cavern of the pit area where the experiment is located.
\end{itemize}

During the Conceptual Feasibility Study~\cite{Arduini:2851946}, the need for an elevator platform providing access to the experiment at all relevant heights has been identified. A feasible solution has been designed by a consultancy firm~\cite{XLIndustries:2023}, which provides a normal operation mode and an evacuation mode complying with the requirements outlined in Sec.~\ref{sec:Overview}.

The main characteristics of this platform are the following:
\begin{itemize}
    \item Total maximum useful load: 500~kg (can be redesigned for up to 1000~kg);
    \item Space for two operators in an adequate protective booth, to avoid risk of personal injuries during fast descent;
	\item Electrical power from a secure network with backup batteries in case of power outage;
    \item Nominal speed in automatic mode 40~m/min;
    \item Nominal speed in manual mode 12~m/min;
	\item Controlled descent in case of evacuation at a speed of 70~m/min to comply with the requirement of a 2-minute maximum descent time for evacuation set out in the Feasibility Study Report~\cite{Arduini:2851946};
    \item Compliance with the EN1495 standard.
\end{itemize}

\begin{figure}[h!]
	\centering
    \includegraphics[width=0.8\textwidth]{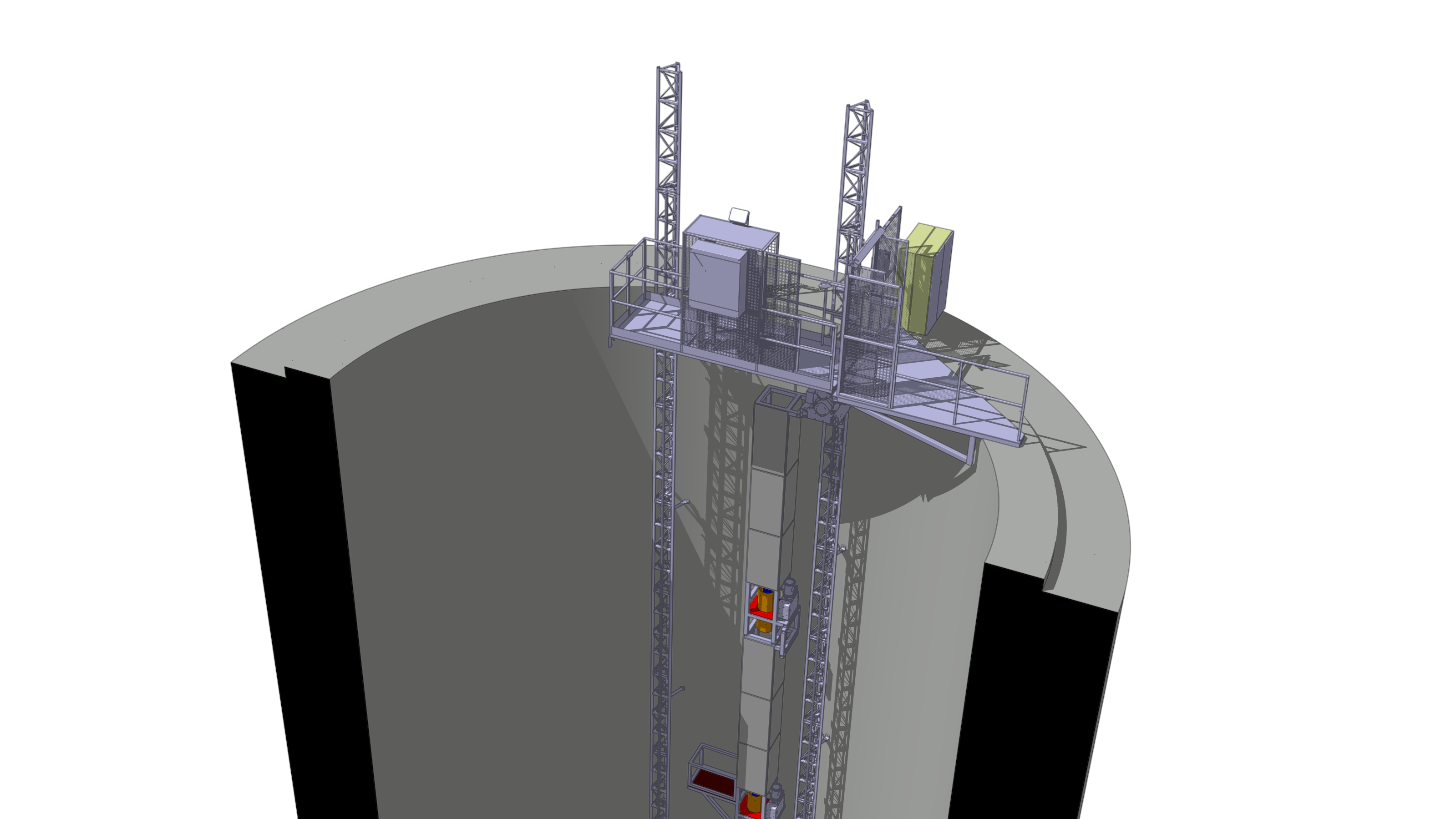}
    \includegraphics[width=0.7\textwidth]{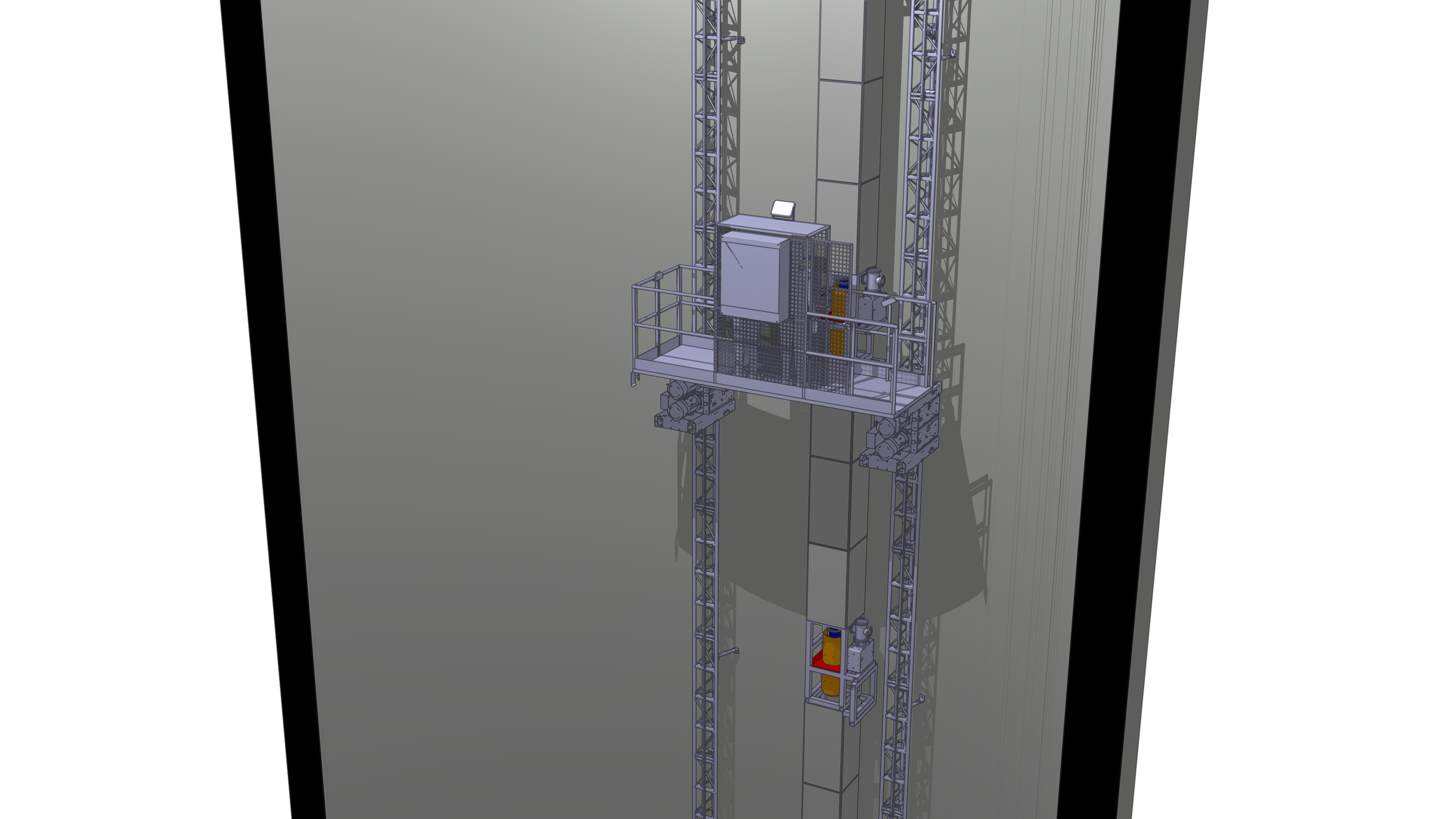}
	\caption{\label{fig:Platform1} {\it From top to bottom}: 3D views of the top landing, and of the dedicated lifting platform in PX46 during operation~\cite{XLIndustries:2023}.}
\end{figure}

The proposed custom-made elevator platform (see Figure~\ref{fig:Platform1}) allows routine access to the experiment, stopping at the locations of the side arms of the experiment via an automatic operation mode like a normal lift. In the event of an emergency evacuation, it could reach the surface with sufficient speed for escaping through the SX4 building and, should evacuation to the surface not be possible, it could descend to the bottom of PX46 with the required speed as specified.

The platform could also be used during the installation of the AI experiment itself. In such a case, a manual operation mode has been envisaged, where the operators could start and stop the platform manually (moving at reduced speed) and thus be able to work at any point along the height of PX46.

An anchor point installed above the platform in the surface building will provide the possibility of rescuing the operators in the event of mechanical failure of the platform itself, the probability of this happening concurrently with a fire or helium release event being considered negligible. In such an event an alternative option could be to suspend a rescue nacelle from the lifting equipment described hereafter, installed on the overhead crane in the SX4 surface building. It could be used autonomously by a Fire Brigade rescue team. The design of the platform should allow enough room and capacity to support the two operators and two rescuers in case of an intervention. 

\begin{figure}[h!]
	\centering
	\includegraphics[width=.5\textwidth]{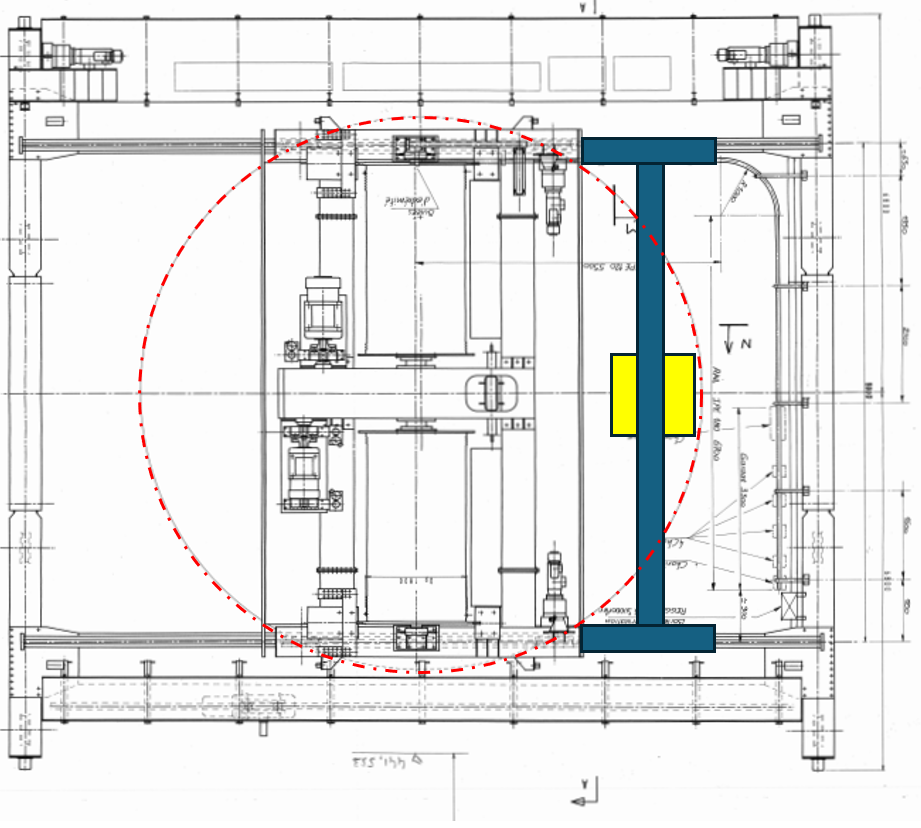}
	\caption{\label{fig:lifting} Top view of the SX4 overhead crane equipped with new lifting equipment dedicated to the AI experiment.}
\end{figure}

The handling of the different elements of the AI experiment during the installation phase will be possible thanks to additional lifting equipment fixed to the main hoist of the SX4 overhead crane (see Figure~\ref{fig:lifting}).

\begin{figure}[h!]
	  \centering
    \includegraphics[width=.7\textwidth]{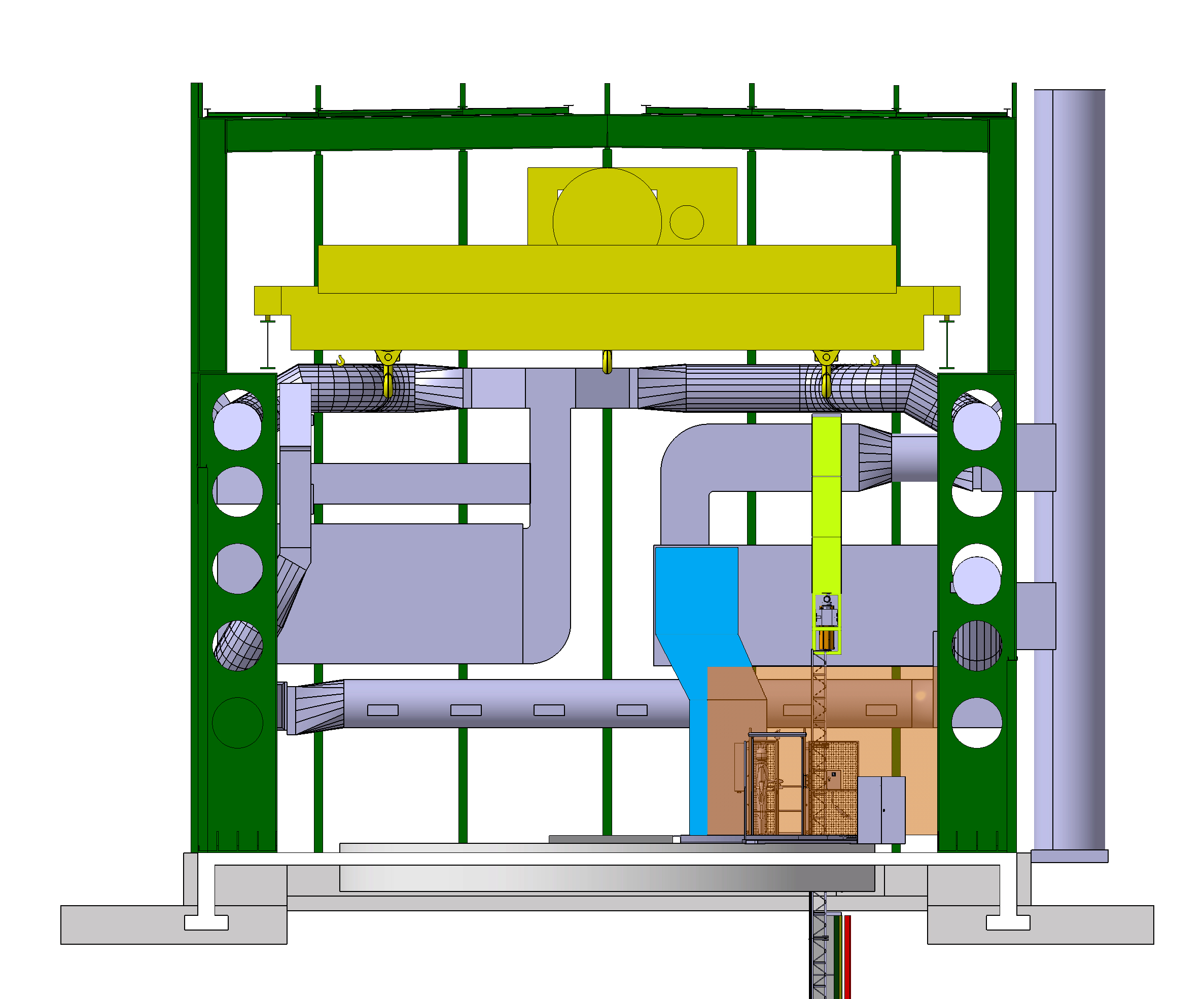}
	\caption{\label{fig:handling} Side view of the SX4 surface building during handling operations : the overhead crane is in yellow, ventilation pipes appear in blue, the AI experiment component in light green, the ventilation room in orange.}
\end{figure}

This solution allows for precise handling in the narrow space between the wall and the access platform over the entire height of the shaft, the cables of the main overhead crane hoist having too wide a span to fit into that volume. These handling operations are illustrated in Figure~\ref{fig:handling}.

\subsection{Mobile shielding wall}

As described in Sec.~\ref{sec:CE}, a shielding wall separation is required for regular access to the bottom of the PX46 shaft, including during LHC beam operations. The requirement for a shielding door with a 4m x 4m opening comes from the need to replace equipment failing in the UX45 cavern within a day, including during LHC operations. 
This system would be completely shielded, including inside the frame. A preliminary design of such system has been proposed in 2025 that demonstrates the feasibility, based on the design of similar equipment implemented recently for HIE-ISOLDE.

%% file: 8-HVAC.tex
\section{Heating, ventilation and air conditioning (HVAC)}
%(O. Crespo-Lopez, A.P. Marion)
\label{sec:HVAC}

%\it 1-2 pages: discuss the possible need of ventilation at the bottom of the shaft upon construction of shielding wall, for limiting CO2 stagnation. Discuss possible impact of the experiment on the PX46 air extraction. Possibly mention chilled water installation system, if performed during LS3. Prepare timeline and cost estimate.

The access doors illustrated in Fig.~\ref{fig:CE:chicane} have a meshed insert. This stems from the requirement of having some airflow at the bottom of PX46 when the shielding wall is constructed in order to avoid stagnation and accumulation of CO2 creating a potential hazard for the AI personnel. Meshed doors comply with this requirement, while not adding any extra impact in case of fire or of an helium release accident, since PX46 and UX45 are already communicating via TU46.

As already mentioned in Sec.~\ref{sec:CE}, an approximately 20~m$^2$ ventilation room (which would eventually expand into the laser laboratory) will be constructed in building SX4 on top of the shaft in order to maintain the slight overpressure inside it (minimum \SI{20}{Pa}, currently \SI{40}{Pa}) guaranteeing unchanged ventilation management of the underground areas. Two cascading doors would the escaping airflow when accessing the room to be minimised.

No further activities will be required in the HVAC domain during the realization of the activities discussed in this report. As mentioned in Sec.~\ref{sec:Overview}, further works can be delayed until the construction phase of the AI.

%\textcolor{red}{Explanation of the ventilation room shown in Fig.~15?}

%% file: 9-OtherServices.tex
\section{Electrical Services}
%(M. Parodi)
\label{sec:EL}

% \it 1-3 pages: description of any extra infrastructure works needed for powering the experiment or the ancillary equipment (elevator shaft, etc.), and the laser lab in SX46, that must be performed during LS3. 

As discussed extensively in \cite{Arduini:2851946} and mentioned in Sec.~\ref{sec:Overview}, no major electrical works are needed to prepare the PX46 site for hosting an AI, since enough electrical power is already available in SX4. The only significant activity would installing the power supply for the elevator. For this, 22~kW are needed from the secured network, with a 63~A feeder. The current switchboard is not fitted for this. It would need to be replaced and consolidated to host a new 63~A feeder with a copper cable \SI{60}{\m} long, 5$\times$25 mm$^2$. Other cabling work and its connection to the main low voltage switchboard can then be delayed to the moment of the construction of the AI and of its laser laboratory.

%% file: 10-CostSchedule.tex
\section{Infrastructure Cost Drivers and Schedule Constraints}
%(All)
\label{sec:Cost}

%\it 1-2 pages: summary of the studies done in previous chapters, possibly integrated into a technical-oriented Gannt chart. Possibly improve cost estimate from Class 5 to Class 4.

The main cost drivers have been discussed in details in the previous Sections, namely the construction of the shielding wall and the large movable shielding door, the elevator platform including modifications of the top plug of PX46 with the ventilation room, and all the necessary modifications to the access control system, the fire detection and evacuation alarms, and the emergency communication devices.  
Cost estimates have been made for all these systems, based on the more detailed engineering analysis performed in this document compared to the initial feasibility study~\cite{Arduini:2851946}. It should be noted that the cost for the cabling activities relevant to the above systems are included in the present cost estimates. The figures are summarized in Table~\ref{tab:total_costs}. These estimates can be considered \textit{Class 4}~\cite{bib:DOEGuide} with a lower range of uncertainty between -15\% and -30\% and an upper range between +20\% and +50\%. Note that we did not include any contingencies in the estimate, however the level of uncertainty at this stage should probably allow enough flexibility.
%\textcolor{red}{Our current estimates are sufficiently detailed to be already in Class 3 in fact, which is what is needed for budget authorization.}

\begin{table}[htbp]
\centering
\caption{Class 4 budget estimates for civil engineering, access and alarm system modifications, lifting platform and other heavy handling and transport equipment to be installed during LS3.}
\smallskip
\begin{tabular}{|l|c|}
\hline
{\bf Description}                                    & {\bf Cost (CHF)} \\
\hline
Civil engineering design services                    & 130'000 \\
Civil engineering works (including concrete blocks and door rails)  & 260'000 \\
LHC Access Safety System (LASS)                      & 65'000 \\
Fire detection, alarms, emergency communications     & 30'000  \\
RP monitor                                           & 20'000  \\
Lifting platform                                     & 400'000  \\
Movable shielding door (structure)                   & 100'000  \\
Modifications of shaft top plug (including ventilation room)                   & 75'000 \\
Additional hoist in SX4                              & 130'000 \\
\hline
{\bf Grand total}                                    & {\bf 1'210'000} \\
\hline
\end{tabular}
\label{tab:total_costs}
\end{table}

We note that the present cost estimate is significantly lower that presented in the initial feasibility study~\cite{Arduini:2851946}. This is due, on the one hand, to a better understanding of the needs and to the more detailed engineering design achieved in the present iteration, resulting in a naturally more precise cost estimate; on the other hand, the initial cost estimate included infrastructure that is pertinent only to the AI experiment, such as a new chilling water production and distribution system and electrical and service cabling systems both for the AI experiment and its laser laboratory, which have not been included in the present study, as not being strictly needed for the purpose of preparing the site during LS3.

\begin{figure}
    \centering
    \includegraphics[width=1.0\linewidth]{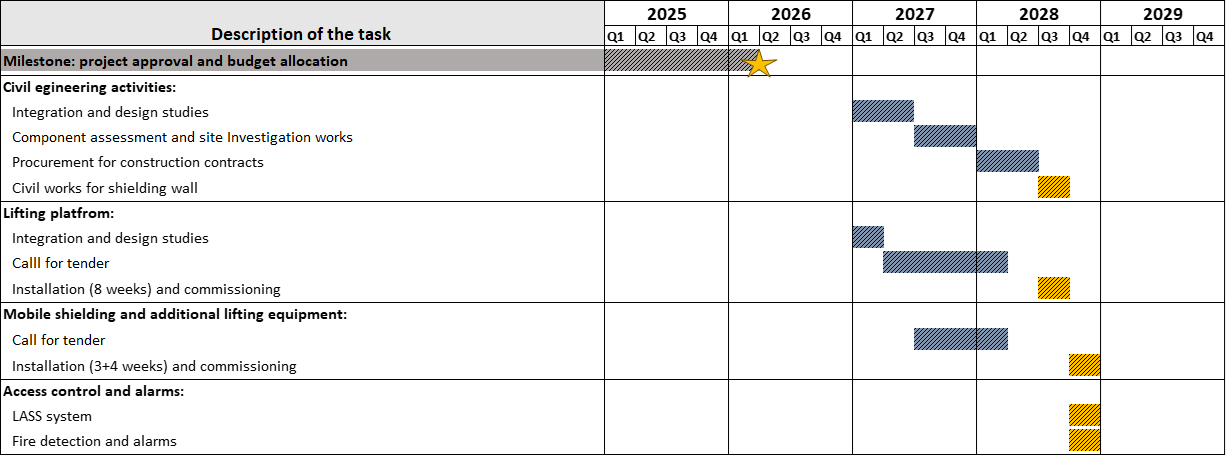}
    \caption{Gantt chart outlining the timeline of the major work items described in this report. Preparatory activities are indicated in gray, while field work is indicated in orange.}
    \label{fig:Gantt}
\end{figure}

A preliminary schedule for the realization of all the works described in this report has been developed, illustrated in Gantt form in Fig.~\ref{fig:Gantt}.
The schedule is subject to a few constraints. Obviously, timely approval of the project and its budget is necessary. Civil engineering works must be completed before installation of the movable shielding door and of the personnel access doors, and before instrumenting them and interfacing with the LASS. The tendering phase of the platform, of the mobile shielding and other lifting equipment has a rather long lead time. Finally, it is not feasible to install fire detection in the shaft before the lifting platform is in place. These constraints are mostly logical and do not actually introduce any significant delays on a technically limited schedule.

The schedule illustrated in Fig.~\ref{fig:Gantt} results from matching the critical-path timeline developed in the present engineering study with the presently (12.08.2025) known draft LS3 schedule.
%illustrated in Fig.~\ref{fig:LS3sched}. 
A potential window of opportunity has been identified starting from August 2027 to end of 2028, in which activities in the LHC Point 4 are limited and would naturally allow for the works pertinent to the present study to be realized~\cite{LS3S:LHC-PM-MS-0022}. This will have to be constantly reviewed in collaboration with the EN/ACE group in charge of the LS3 planning. At present, the relevant technical services have specific internal time frames for the installation work, preferentially in the second half of 2028.
An earlier implementation of the works relevant to this project might also be technically feasible, depending on the approval time and on the available funding and resources.

%\begin{figure}
%    \centering
%    \includegraphics[width=1\linewidth]{Figures/LS3 schedule.png}
%    \caption{Preliminary LS3 schedule, yet to be finalised and approved by CERN management~\cite{LS3S:LHC-PM-MS-0022}. The time slot having the lowest risk of interference with other activities in Point 4, from August 2027 to December 2028, is indicated in pink.}
%    \label{fig:LS3sched}
%\end{figure}

We underline that it is beyond the scope of this document to discuss the availability of the resources to perform the described activities. This is to be decided by CERN management upon approval of this project. Nevertheless, a preliminary informal assessment suggests that the required coordination effort by CERN staff is within very reasonable limits, while most of the activities can be outsourced to contractors.

%% file: 11-Conclusions.tex
\section{Conclusions}
%(G. Arduini, S. Calatroni, O. Buchmuller, J. Ellis)
\label{sec:Conclusions}

Building upon the previous feasibility study~\cite{Arduini:2851946}, this implementation study confirms that the PX46 shaft at LHC Point 4 can be adapted during LS3 (June 2026 - June 2030) to host the installation and operation of a vertical Atom Interferometer during Run 4 without impeding HL-LHC operations. Detailed civil-engineering modifications, bespoke radiation shielding, access-control and alarm systems, and a mobile elevator platform have all been scoped, costed and scheduled. Our refined Class 4 estimate (1.2~MCF $\pm$ 30\%/50\%) and definition of a critical-path timeline (1.5 years from formal approval) demonstrate that---with timely authorisation---all interventions are achievable within the LS3 window.

The technical assessment reveals no fundamental barriers to implementation. The radiation protection analysis validates the proposal in~\cite{Arduini:2851946} of shielding measures that will reduce PX46 to a simple Supervised Radiation Area, enabling safe concurrent operation with HL-LHC. The proposed engineering solutions for personnel access, emergency evacuation, and equipment handling represent established approaches adapted to the unique LHC environment.

The strategic approach of decoupling infrastructure preparation from experiment construction offers significant operational advantages, allowing subsequent AI installation and commissioning to proceed in parallel with HL-LHC operations. The physics case remains compelling, with unique sensitivity to ultra-light dark matter and pioneering gravitational wave detection capabilities in the intermediate frequency range.

These findings provide the technical foundation necessary for informed decision-making regarding project authorisation, resource allocation, and integration with the broader LS3 schedule and HL-LHC programme objectives.

%{\it 1 page: summary and outlook.}